\documentclass[%
 reprint,
superscriptaddress,
 amsmath,amssymb,
 aps,
pra,
]{revtex4-2}

\usepackage{graphicx}
\usepackage{dcolumn}
\usepackage{bm}
\usepackage{babel}

\usepackage{comment}
\usepackage{physics}
\usepackage{footnote}
\usepackage{subcaption}
\usepackage[abs]{overpic}
\usepackage{booktabs}
\usepackage{amssymb}
\usepackage{hyperref}
\usepackage{multirow}

\usepackage{xcolor}

\usepackage[normalem]{ulem} 
\usepackage{color}
\usepackage[dvipsnames]{xcolor}

\begin{document}

\preprint{APS/123-QED}

\title{Distributed Architectures for Quantum Extreme Learning Machines}
\title{Learning functions of quantum states with distributed architectures}

\date{\today}
\author{
    Marta Gili$^{1,2}$, Eliana Fiorelli$^{2}$, Ane Blázquez-García$^{1}$, Gian Luca Giorgi$^{2}$, Roberta Zambrini$^{2}$ \\
    {\small $^{1}$Ikerlan Technology Research Centre, Basque Research and Technology Alliance (BRTA), J.M. Arizmediarrieta Pasealekua 2, Arrasate-Mondragón, 20500 Arrasate-Mondragón, Spain} \\
    {\small $^{2}$Institute for Cross-Disciplinary Physics and Complex Systems (IFISC), UIB-CSIC, Campus Universitat Illes Balears, 07122 Palma, Spain} \\
}

\begin{abstract}

Distributed architectures are gaining prominence in quantum machine learning as a means to overcome hardware limitations and enable scalable quantum information processing. In this context, we analyze the design and performance of distributed Quantum Extreme Learning Machine (QELM) architectures for learning functions of quantum states directly from data, restricting measurements to easily implementable projective measurements in the computational basis. The aim is to determine which schemes can effectively recover specific properties of input quantum states, including both linear and nonlinear features, while also quantifying the resource requirements in terms of measurements and reservoir dimensionality. 
We compare standard three-layer QELM  with a spatially multiplexed architecture composed of multiple independent three-layer units for linear (quantum) tasks, showing a linear reduction in resource requirements per unit.
For nonlinear properties, the study examines the multiple-injection architecture and introduces a novel distributed design that incorporates entanglement between subsystems within a spatially multiplexed framework, evaluating its performance through the reconstruction of complex nonlinear quantities such as polynomial targets, Rényi entropy, and entanglement measures. Our results demonstrate that the distributed design enables the reconstruction of higher-order nonlinearities by increasing the number of interacting subsystems with reduced resources, rather than increasing the size of an individual reservoir, providing a scalable and hardware-efficient route to quantum property learning.


\end{abstract}


\maketitle


\section{\label{sec:int}Introduction}

Distributed quantum computing has emerged as a promising approach to overcome the limitations of Noisy Intermediate-Scale Quantum (NISQ) devices by interconnecting multiple quantum processors through quantum networks, enabling larger and more complex computations than a single processor can handle \cite{boschero, caleffi_distributed_2024, barral}. In parallel, distributed quantum machine learning has gained significant attention as an effective paradigm to leverage distributed quantum computing for quantum-enhanced learning tasks. Current approaches include quantum federated learning, which aggregates locally trained models to preserve data locality and privacy \cite{Chen, Ren}, and model-parallel distributed quantum neural networks, which reduce per-device qubit requirement by partitioning circuits or ensembling smaller quantum neural networks \cite{Wu,marshall2023high}. Other strategies rely on local operations with classical communication \cite{hwang_distributed_2024}, linking parameterized circuits through measurement and feed-forward, while others exploit shared entanglement to distribute computation across quantum processing units using nonlocal operations or state teleportation \cite{Khait}. Measurement-based photonic architectures for distributed quantum machine learning based on continuous-variable cluster states have also been proposed recently \cite{garcia-beni_quantum_2025}.

In this work, we address the challenge of scalable quantum learning of (functions of) quantum states by introducing and analyzing distributed architectures. We consider a reservoir-based approach known as Quantum Extreme Learning Machine (QELM) \cite{MujalEtAl21}. This approach overcomes challenges of training parameterized quantum circuits, such as barren plateaus and high computational costs \cite{mcclean2018barren}, avoiding the need for iterative optimization and backpropagation. It relies on fixed quantum reservoirs that transform input data into a high-dimensional feature space, followed by minimal classical post-processing \cite{xiong_fundamental_2024} that requires only a simple training step (e.g., linear regression) on the measurement outcomes, making them particularly attractive for near-term quantum hardware \cite{suprano_experimental_2024}. Since they operate on quantum substrates, QELMs are naturally suited for processing quantum states as reported in quantum state classification, entanglement detection, and quantum state reconstruction \cite{Ghosh2019,innocenti_potential_2023, zia_quantum_2025}. Furthermore, reservoir-based methods and feature maps have been recently explored for classical data processing, introducing spatial multiplexing \cite{nakajima_boosting_2019, paparelle}, interlinked modular quantum architectures \cite{higher,lau_modular_2024}, feature maps in parallel scenarios \cite{Matsumoto} and cluster states \cite{garcia-beni_quantum_2025}. 
 
Extending these advances to distributed architectures operating on $quantum$ data and tasks presents an important and timely direction. To this end, we introduce a distributed QELM framework for learning and reconstructing properties of quantum states. Overall, our contribution lies in a systematic study of distributed QELMs with three primary objectives: (i) {\it resource scaling,} quantifying the required resources, such as measurement counts and reservoir dimensions, across different distributed architectures; (ii) {\it performance benchmarking,} assessing whether and under which conditions distributed strategies provide tangible advantages over centralized approaches; and (iii) {\it architectural mapping,} elucidating the relationship between architectural design choices and the complexity of the accessible quantum properties, encompassing both linear observables and nonlinear functions of the quantum state.

The rest of the paper is structured as follows. In Section \ref{sec:background}, we introduce the theoretical background for QELMs and explain how they can be used to reconstruct properties of quantum states. Building on this foundation, Section \ref{sec:architectures} presents the architectures under analysis. Next, Section \ref{sec:scaling} compares these architectures in terms of the quantum resources they require. Section \ref{sec:results} then presents the results obtained from applying the architectures to the reconstruction of various linear and nonlinear properties. Finally, Section \ref{sec:conc} discusses conclusions and outlines directions for future work.

\section{Background}
\label{sec:background}

\subsection{Quantum Extreme Learning Machines}

An Extreme Learning Machine (ELM) is a supervised learning technique designed to solve static machine learning tasks \cite{HUANG2006489, huang2011extreme}.  It uses a fixed, generally unoptimized classical substrate as a single hidden-layer feedforward network to process data. More generally, a system with complex dynamics can be used to map input signals $\bm{x}$ into their state space, $\bm{f}(\bm{x})$. Then, solving a given task requires only a simple training step at the output, which is typically a linear regression. More formally, considering a target vector $\bm{y}$ and a training dataset $\left\lbrace (\bm{x}^{\mathrm{tr}}_i,\bm{y}_i^{\mathrm{tr}})\right\rbrace _{i=1}^{N_{\mathrm{tr}}}$, where $N_{\mathrm{tr}}$ is the number of samples, the supervised learning procedure involves a linear regression to fix a linear map, $W$, by minimizing the distance between the true target $\bm{y}_{i}^{\mathrm{tr}}$, and the predicted one, $W(\bm{f}(\bm{x}^{\mathrm{tr}}_{i}))$.

In QELM, the classical substrate is replaced by a quantum system. This enables the direct processing of quantum-state properties, circumventing the need for prior quantum state embedding \cite{MujalEtAl21}. The training dataset is given by $\left\lbrace (\rho^{\mathrm{tr}}_i, \bm{y}_i^{\mathrm{tr}}) \right\rbrace_{i=1}^{N_{\mathrm{tr}}}$, where $\rho^{\mathrm{tr}}_i$ represents the input quantum state and $\bm{y}_i^{\mathrm{tr}}=\{(y_i^{\mathrm{tr}})_l\}_{l=1}^{N_{\mathrm{tg}}}$ is the corresponding target vector, with $(y_i^{\mathrm{tr}})_l$ representing a specific target and $N_{\mathrm{tg}}$ the total number of targets to be estimated. For simplicity, we will use $\rho$ and $\bm{y}=\{y_l\}_{l=1}^{N_{\mathrm{tg}}}$ to denote generic quantum states and target vectors. Input states $\rho \in \mathcal{H}_S$ are processed via interaction with reservoir $R$ states $\eta \in \mathcal{H}_R$. The evolution of the global system $\Psi= S \cup R$, with Hilbert space $\mathcal{H}_{\Psi} = \mathcal{H}_S \otimes \mathcal{H}_R$, is governed by a Completely Positive and Trace-Preserving (CPTP) quantum map, $\Gamma: \mathcal{H}_{\Psi} \longrightarrow \mathcal{H}_{\Psi} $. The output layer can be obtained by performing a Positive Operator-Valued Measure (POVM) measurement on the global state. While POVMs enable fundamentally optimal performance in certain tasks, these measurements are experimentally demanding, requiring more complex implementation (e.g, with ancillae) and resources. A simpler approach is to restrict to Projection-Valued Measures (PVMs) in one basis, which can be selected for ease of implementation. Consequently, the output is given by the linear map or functional
\begin{equation}\label{e_qrp_traing}
    y_l = \sum_{k=1}^{p} W_{lk} \mathrm{Tr}[ E_k \Gamma(\rho \otimes \eta) ] \, ,
\end{equation}
with $\left\lbrace E_{k}\right\rbrace_{k=1}^p$   PVM on the global state and $p$ number of possible outcomes \cite{innocenti_potential_2023}.

The training procedure aims to optimize the weights $\bm{W} \in \mathbb{R}^{N_{\mathrm{tg}} \times p}$ in order to minimize the distance between each target $ (y_i^{\mathrm{tr}})_{l}$ and the corresponding output layer $ \sum_{k} W_{lk} \mathrm{Tr}[ E_k \Gamma(\rho^{\mathrm{tr}}_i \otimes \eta) ] $, $\forall i \in \{1,..., N_{\mathrm{tr}}\}$. This is achieved via a linear regression step.
Thus, the building blocks of this framework are the three layers illustrated in Figure \ref{fig:architectures}(a): an input state, a CPTP quantum map, and a set of measurements.

\subsection{Quantum Property Reconstruction}

\begin{figure*}[t]
    \centering

    \vspace{0.6cm}
    
    \begin{subfigure}[c]{0.35\linewidth}
        \centering
        \begin{overpic}[width=\linewidth]{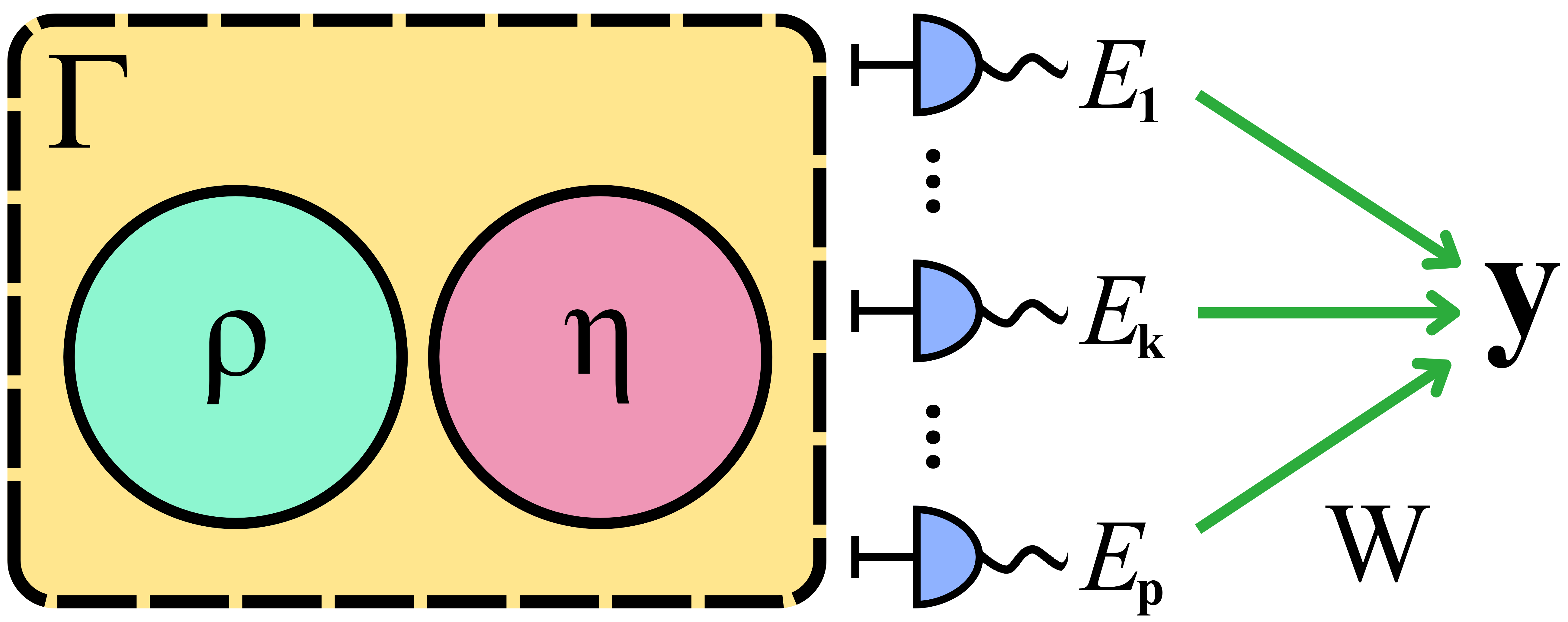}
            \put(2,78){\textbf{(a) Single three-layer architecture (S3L)}}
        \end{overpic}
        \label{fig:arch_a}
    \end{subfigure}
    \hspace{0.05\linewidth}
    \begin{subfigure}[c]{0.55\linewidth}
        \centering
        \begin{overpic}[width=1.05\linewidth]{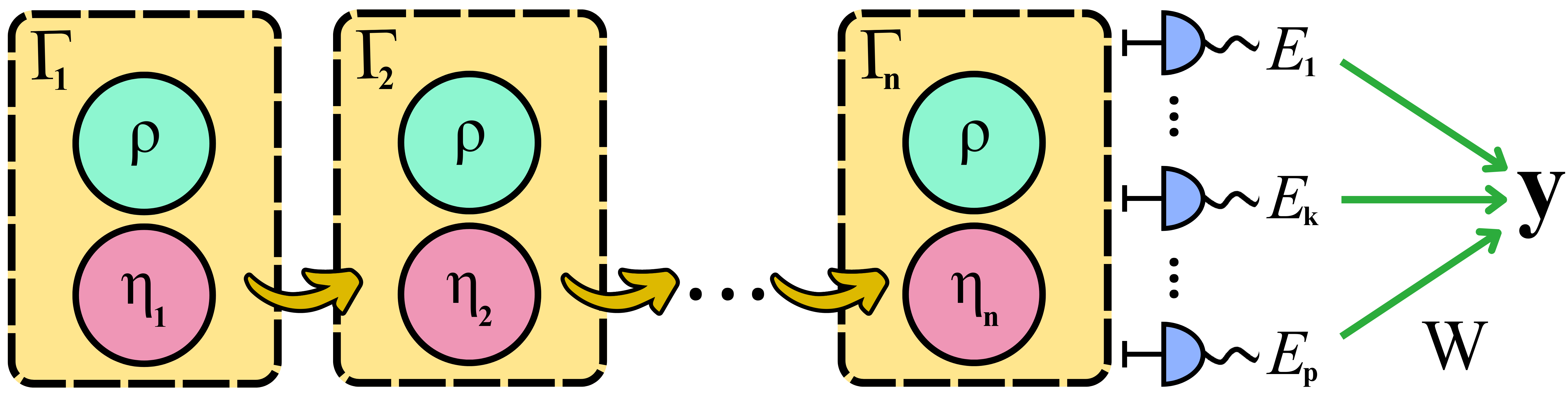}
            \put(2,81){\textbf{(c) Multiple-injections architecture (MI)}}
        \end{overpic}
        \label{fig:arch_c}
    \end{subfigure}

    \vspace{0.6cm}

    \begin{subfigure}[c]{0.45\linewidth}
        \centering
        \begin{overpic}[width=0.98\linewidth]{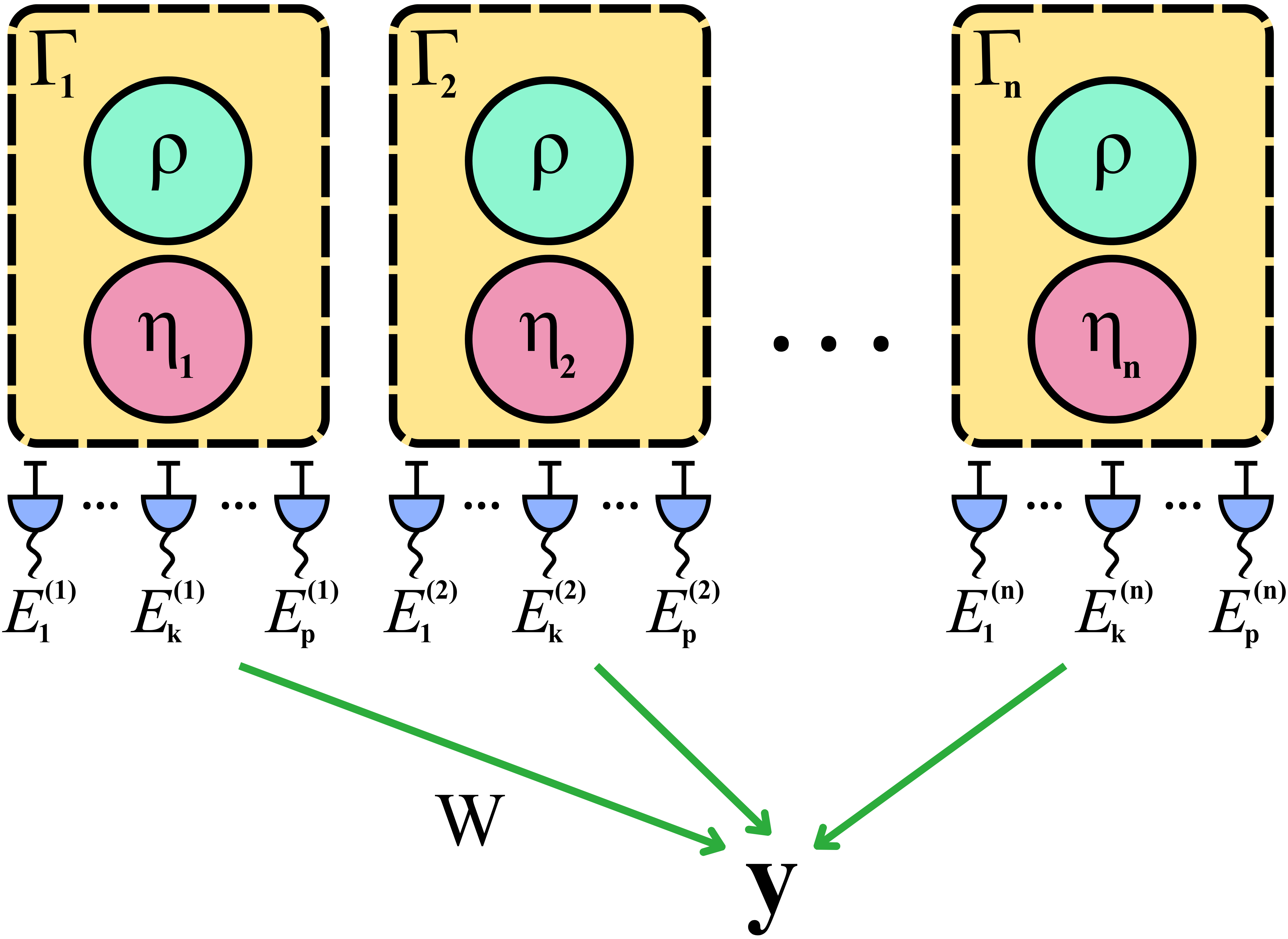}
            \put(2,174){\textbf{(b) Spatially multiplexed architecture (SM)}}
        \end{overpic}
        \label{fig:arch_b}
    \end{subfigure}
    \hspace{0.04\linewidth}
    \begin{subfigure}[c]{0.45\linewidth}
        \centering
        \begin{overpic}[width=1.021\linewidth]{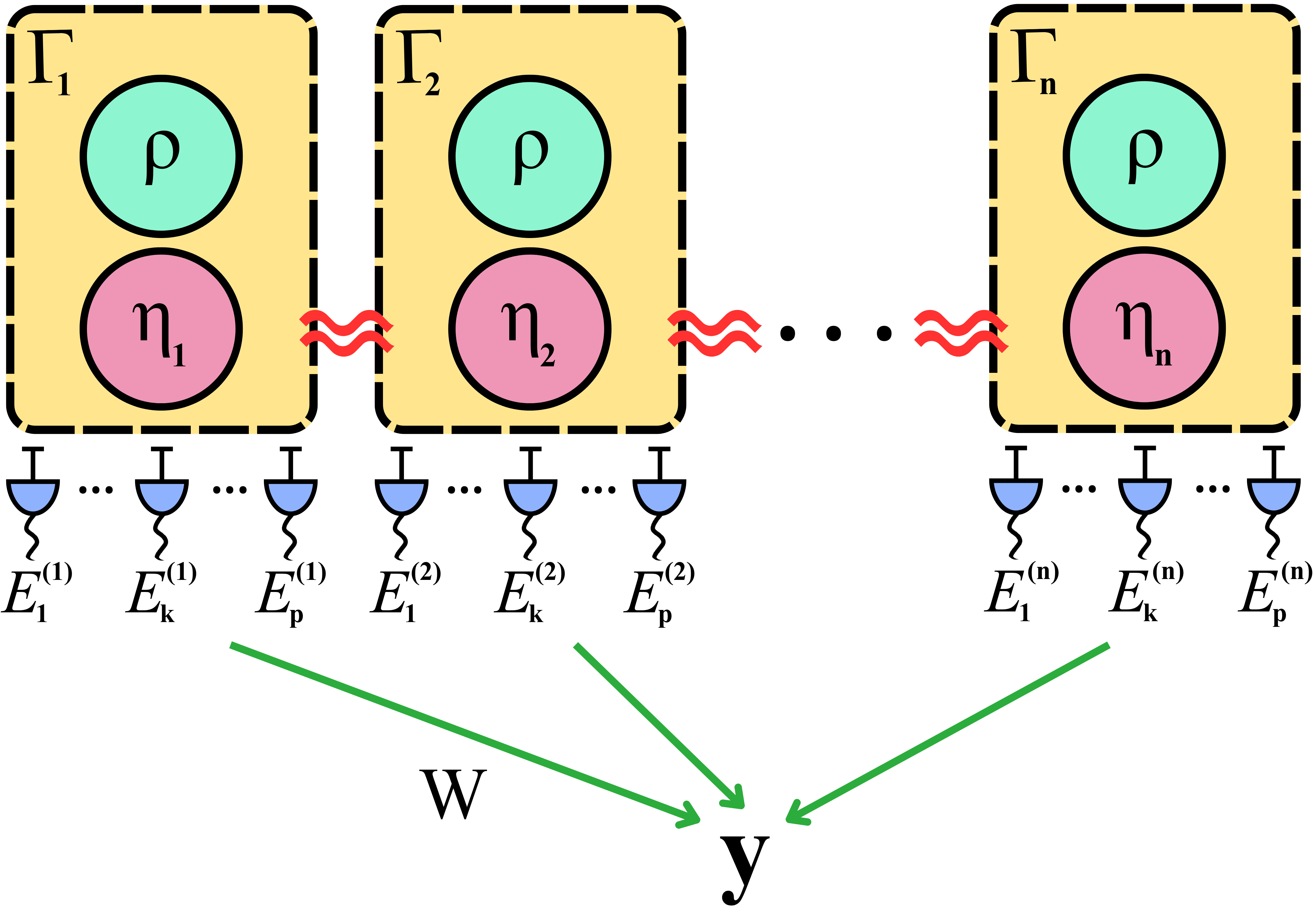}
            \put(2,170){\textbf{(d) Distributed architecture (D)}}
        \end{overpic}
        \label{fig:arch_d}
    \end{subfigure}

    \caption{Schematic setup of the architectures. On the left are the architectures for reconstructing linear properties: (a) and (b). On the right are the architectures for reconstructing nonlinear properties: (c) and (d).}
    \label{fig:architectures}
\end{figure*}

This work focuses on applying QELM architectures to quantum property reconstruction. The starting observation is that in general a PVM measurement in a single basis is easy to implement but does not provide enough information for full quantum state tomography.
On the other hand, the three-layer structure of Figure \ref{fig:architectures}(a) enables access to properties of the input state $\rho$, since the input–output relation in Eq.~\eqref{e_qrp_traing}  corresponds to reconstructing observables on  $\rho$ via a collection of effective measurements \cite{innocenti_potential_2023}. 
In particular, considering the target as the input state, PVMs on the larger Hilbert space are expected to provide a faithful realization of the corresponding POVM by Naimark dilation \cite{Busch2016QuantumMeasurement}. Equivalently, the overall process of measuring the global system state after the evolution $\Gamma$ can be reframed as an effective measurement performed directly on the input $\rho$:
\begin{equation}\label{e.taskQELM}
    y_{l} = \sum_k W_{lk} \mathrm{Tr}_S [ \Tilde{E}_k \rho ] ,
\end{equation}
where 
\begin{equation}
\Tilde{E}_k = \mathrm{Tr}_R [\Gamma^{\dagger}[E_k] (\mathbb{I}_S \otimes \eta) ]
\end{equation}
represents an effective POVM on the state $\rho$ and $\Gamma^{\dagger}$ is the adjoint of the CPTP map $\Gamma$. Furthermore, this can be rewritten as
\begin{equation}\label{e.taskQELM_1}
    y_{l} = \mathrm{Tr}_S \left[ O_l \rho\right] ,
\end{equation}
where $O_l = \sum_k W_{lk} \Tilde{E}_k $ is the reconstructed observable on the input system $S$. For the reconstruction of arbitrary operators $O_l$ (i.e., accessing the complete tomography of $\rho$), some requirements need to be met.

First, assuming non-entangling measurements between the input and the reservoir, the CPTP map $\Gamma$ has to generate correlations between the input and the reservoir to attain any enhancing effect from the reservoir when reconstructing an arbitrary input. Otherwise, no information about the input is transferred to the reservoir, making it impossible to extract useful information from it. This is further explored in Appendix \ref{sec:dynamics}. Additionally, the set of effective POVMs $\lbrace \Tilde{E}_k \rbrace_{k=1}^{p}$ must be informationally complete. This means that it has to span the space of Hermitian operators on $\mathcal{H}_S$ (i.e., $O_l \in \mathrm{span}_{\mathbb{R}}(\lbrace \Tilde{E}_k \rbrace)$). By knowing the dimension of this space, we can determine the minimum number of independent measurement outcomes required for full reconstruction.

For a single qubit, the density matrix is fully specified by the three real components of the Bloch vector, so at least three independent measurement outcomes are required to reconstruct the state. A single-basis PVM is insufficient for full quantum state tomography. Although a Symmetric, Informationally Complete Positive Operator-Valued Measure (SIC-POVM) enables complete reconstruction from one measurement setting \cite{sic}, it is experimentally demanding on current hardware \cite{exp_sic, exp_sic2}. Standard tomography instead uses PVMs in three distinct bases (e.g., $\Pi_x,\Pi_y,\Pi_z$), which is simpler but still costly. In QELM, the qubit is effectively embedded in a larger Hilbert space. A wider set of outcomes is accessed via easier measurements, e.g., a single global PVM in the computational basis ($\Pi_z$ on the joint system–reservoir). By Naimark’s theorem, any POVM can be implemented as a PVM in an enlarged space after a suitable global unitary with an ancilla \cite{Oszmaniec}.

%

For this purpose, we need the dimension of the reservoir to be such that the resulting joint Hilbert space is sufficiently large and we consider $\lbrace {E_{k}} \rbrace = \{\ket{j}\bra{j}\}_{j=0}^{d-1}$, where $d$ denotes the dimension of the global Hilbert space $\mathcal{H}_S \otimes \mathcal{H}_R$. This choice of PVMs sets the number of outcomes to $p=d$, which can be adjusted depending on the number of qubits that we introduce in the reservoir.


Quantum state properties can be either linear in the quantum state, such as observables or the state itself (in tomography), which can be addressed as for Eqs.~\eqref{e.taskQELM} and \eqref{e.taskQELM_1}, or nonlinear, such as purity or entanglement measures that depend on higher-order functions of the state. In the following, we will address
both cases.
     
\section{Architectures}
\label{sec:architectures}

Different architectures can be designed and compared by benchmarking their performance on various quantum state tasks while optimizing resource utilization. In this section, we present four such architectures, for which we derive conditions on the number of measurement outcomes required for accurate property estimation and determine the corresponding reservoir dimension needed to achieve it. We anticipate that the complexity of the target tasks varies depending on the type of architecture, and we will address not only those designed for linear properties (observable expectation on the input state) but also for nonlinear property reconstruction.

\subsection{Single three-Layer Architecture}

The first architecture under consideration consists of a single three-layer structure, which corresponds to the standard QELM model (Figure \ref{fig:architectures}(a)). Hereafter, we refer to this architecture as the single three-layer (S3L) architecture. In this configuration, the output is given by Eq.~\eqref{e_qrp_traing}, which defines a linear mapping between the input state $\rho$ and the measured expectation values. As a consequence, this architecture is fundamentally limited to reconstructing linear functionals of the input state, such as expectation values of observables. This restriction is a direct consequence of the linearity of quantum operations and measurements acting on a single copy of $\rho$, and it prevents access to intrinsically nonlinear properties, including higher-order moments, entropic quantities, and entanglement measures. Overcoming this limitation requires architectural extensions that effectively enlarge the space on which the input state is processed, a strategy that motivates the alternative designs introduced in the following sections.

To enable the reconstruction of arbitrary observables (i.e., full state reconstruction), the condition for the set of effective measurements to be informationally complete needs to be met \cite{innocenti_potential_2023}. In this case, the space of Hermitian operators on $\mathcal{H}_S$ has dimension $\dim(\mathcal{H}_S)^2$, meaning that the number of outcomes must satisfy $p \geq \dim(\mathcal{H}_S)^2$, with equality holding when the set $\lbrace \Tilde{E}_k \rbrace$ consists of linearly independent operators. However, since we are dealing with the reconstruction of density matrices, we must also account for the constraint $\mathrm{Tr} \left( \rho \right) = 1$. This removes one degree of freedom, meaning that, in practice, $p \geq \dim(\mathcal{H}_S)^2-1$ is required. 


For this architecture, the number of possible outcomes using the PVMs in the computational basis is $p = \dim(\mathcal{H}_S)\dim(\mathcal{H}_R)$. Notably, due to the completeness relation, the number of independent measurements is $p-1$. As a result, a necessary condition for full input reconstruction is 
\begin{equation}\label{eq:single}
    \dim({\mathcal{H}_R}) \geq \dim({\mathcal{H}_S})
\end{equation}
This means that under the established measurement settings, full reconstruction of the quantum state is achievable if the reservoir’s Hilbert space dimension is no smaller than that of the system.
  


\subsection{Spatially Multiplexed Architecture}
We now consider the spatially multiplexed (SM) architecture, where the same input state is injected simultaneously into multiple independent reservoirs, enabling parallel processing without inter-unit communication. Architectures with spatial multiplexing have shown promising results \cite{nakajima_boosting_2019}, including in experimental implementations \cite{paparelle}. As represented in Figure \ref{fig:architectures}(b), multiple non-interacting three-layer structures---hereafter referred to as units--- act in parallel, with each output layer contributing to the training step. This approach effectively increases the dimensionality of the feature space by distributing the computation across multiple independent quantum subsystems, avoiding the need to scale up the size of individual quantum processors and making it particularly interesting for near-term quantum hardware.

More in detail, $n$ copies of the input system $S$ are prepared in a state $\rho$, and each one is let to interact with the $i$-th reservoir in the state $\eta_i$. The state $\rho \otimes \eta_i $ is evolved by means of a CPTP map $\Gamma_i: \mathcal{H}_{S}\otimes \mathcal{H}_{R_i} \rightarrow \mathcal{H}_{S}\otimes \mathcal{H}_{R_i}$, for $i = 1,...,n$. Then, the output layer is reconstructed by performing the same set of PVMs on each unit (i.e., $E_k^{(i)}=E_k$ $\forall i \in \{1,...,n\}$). Thus, the QELM output reads
\begin{equation}
\begin{split}
y_l &= \sum_i \sum_k W_{lk}^{(i)}  \mathrm{Tr} [E_k \Gamma_i(\rho \otimes \eta_i)] \\
&=  \sum_{i} \sum_k W_{lk}^{(i)} \mathrm{Tr}_S [\Tilde{E}_k\rho] 
= \mathrm{Tr}_S[O_l \rho] \, ,
\end{split}
\end{equation} 
where $\Tilde{E}_k = \mathrm{Tr}_{R_i} [ \Gamma^{\dagger}_i[E_k](\mathbb{I}_S \otimes \eta_i)]$. As before, it is possible to reconstruct any linear functional $O_l$ of the state $\rho$ under some conditions. In order to avoid redundancies, the CPTP maps $\Gamma_i$ must not only generate correlations but also differ from one another. This requirement is a design choice, as comparable results could be obtained by using identical maps together with different measurements in each unit. Under this setting, the condition for informational completeness reads 
\begin{equation}
\sum_{i=1}^{n} \left( \dim(\mathcal{H}_{R_i})  \dim(\mathcal{H}_{S}) - 1 \right) \geq \dim(\mathcal{H}_{S})^{2} - 1 \, ,
\end{equation}
since $p \geq \dim(\mathcal{H}_S)^2-1$ it is still required, but the number of independent outcomes is now given by the sum over all units. When taking reservoirs of the same dimension (i.e., $ \dim(\mathcal{H}_{R_i}) \equiv \dim(\mathcal{H}_{R})$ $\forall i$) it is 
\begin{equation}
\dim(\mathcal{H}_{R}) \geq \frac{1}{n} \left( \mathrm{dim}(\mathcal{H}_{S}) + \frac{n-1}{\mathrm{dim}(\mathcal{H}_{S})}\right).
\label{eq:sm}
\end{equation}
Therefore, the dimension of each reservoir depends on the total number of reservoirs considered. A resource comparison with Eq.~\eqref{eq:single} is discussed in Section \ref{sec:scaling}.


\subsection{Multiple-Injections Architecture}

The multiple-injections architecture (MI) is conceptually related to time multiplexing, as it exploits multiple sequential injections of the input system into the same quantum reservoir to effectively enhance its expressivity \cite{innocenti_potential_2023}. The corresponding scheme is illustrated in Figure \ref{fig:architectures}(c).

It operates by preparing $n$ copies of the system state $\rho$ while initializing a single reservoir in a state $\eta_0$. The CPTP map now takes states from $\mathcal{H}_{S} \otimes \mathcal{H}_{R}$ to $ \mathcal{H}_{R}$ (i.e., $\Tilde{\Gamma}: \mathcal{H}_{S}\otimes \mathcal{H}_{R} \longrightarrow \mathcal{H}_{R}$), such that the $i$-th reservoir state can be written as $\eta_i = \Tilde{\Gamma}(\rho \otimes \eta_{i-1})$, for $i=1,...,n-1$. For instance, $\Tilde{\Gamma}$ can be thought of as the map for the reduced state of the reservoir in QRC based on erase and write dynamics \cite{FujiN17}, in which case the following relation holds
\begin{equation}
    \Tilde{\Gamma}(\rho \otimes \eta) = \mathrm{Tr}_S [\Gamma(\rho \otimes \eta)],
\end{equation}
being $\Gamma$ the CPTP map coupling the input state and reservoir. To enable the use of global PVMs, we assume that the last introduced input state is not traced out (i.e., the final measurement is over $S \cup R$), so that the measured global system is $\Gamma(\rho \otimes \eta_{n-1})$. Importantly, CPTP maps are associative and can be composed, meaning that multiple sequential applications of the same map can be rewritten as a single equivalent transformation. The final state of the reservoir is the result of a single global map $\Lambda$ acting on all $n$ input system copies and the initial reservoir state:
\begin{equation}
    \eta_{n} = \Lambda(\rho^{\otimes n} \otimes \eta_0).
\end{equation}
$\Lambda$ is itself a CPTP map that encapsulates the entire transformation, without iterative applications. 

In this framework, the output can be rewritten in terms of effective measurements as
\begin{equation}
\begin{split}
    y_l &= \sum_k W_{lk}  \mathrm{Tr} [E_k \Lambda(\rho^{\otimes n} \otimes \eta_0)] \\
    &= \sum_{k} W_{lk} \mathrm{Tr}_S [\Tilde{E}_k \rho^{\otimes n}] =  \mathrm{Tr}_S [O_l \rho^{\otimes n}], 
\end{split}
\end{equation}
where $\Tilde{E}_k = \mathrm{Tr}_{R} [ \Lambda^{\dagger}[E_k](\mathbb{I}_S^{\otimes n} \otimes \eta_0)]$. This result shows that the approach enables the reconstruction of nonlinear functionals of the input state, whose nonlinearity depends on the number of injections $n$ that are performed.

Note that the condition for the set of effective measurements to be informationally complete now requires that they span the space of Hermitian operators on $\mathcal{H}_S^{\otimes n}$. Denoting $\dim({\mathcal{H}_S}) = s$, this space has dimension
\begin{equation}
    d_{s n} = \binom{s^2-1+n}{n},
\end{equation}
which implies that the required number of independent outcomes becomes $p \geq d_{sn} - 1$. This corresponds to the number of independent parameters (degrees of freedom) of a state constrained to the symmetric subspace of $\mathcal{H}_S^{\otimes n}$. The symmetry arises because each injection of the input state is indistinguishable. Therefore, considering that the number of independent PVM outcomes is $p = \dim(\mathcal{H}_S)\dim(\mathcal{H}_R)-1$, the condition on the reservoir dimension to enable the reconstruction of any nonlinear functional of the state $\rho$ is given by
\begin{equation}
    \dim(\mathcal{H}_R) \geq \frac{d_{sn}}{\dim(\mathcal{H}_S)} .
    \label{eq:mi}
\end{equation}
Now the required reservoir dimension depends not only on the dimension of the input system, but also on the number of injections through $d_{sn}$.
 
\subsection{Distributed Architecture}

Finally, we present an alternative architecture that enables the reconstruction of nonlinear properties based on a distributed design, depicted in Figure \ref{fig:architectures}(d). This proposal, referred to as distributed (D) architecture, combines the concept of spatial multiplexing with the introduction of entanglement between reservoirs in order to achieve nonlinearities.

As in the spatially multiplexed architecture, it considers $n$ 3-layer structures, with each input system $S$ prepared in the same state $\rho$, and each reservoir $R_{i}$ initialized in a state $\eta_i$, evolving with $\Gamma_i: \mathcal{H}_{S}\otimes \mathcal{H}_{{R}_{i}} \longrightarrow \mathcal{H}_{S}\otimes \mathcal{H}_{{R}_{i}}$. Additionally, the whole system $\cup_{i=1}^{n} S \cup R_i$ further evolves according to a second CPTP map, $\Phi: \otimes_{i=1}^{n} \mathcal{H}_{S} \otimes \mathcal{H}_{R_{i}} \longrightarrow \otimes_{i=1}^{n} \mathcal{H}_{S} \otimes\mathcal{H}_{R_{i}}$, which is responsible for generating entanglement. After the evolution, the set of measurements $\lbrace {E_{k} \rbrace}$ acts on the whole $\bigcup_{i=1}^n S \cup R_i$. In practice, however, since we restrict ourselves to local measurements on each unit, the corresponding global expectation values can be reconstructed experimentally by combining the local measurement outcomes through classical communication \cite{alves2003direct}.

Taking the latter into account, the output layer is given by

\begin{align}
        y_l = \sum_{k} W_{lk} \,
    \mathrm{Tr} \!\Bigg[
        E_k \,
        \Phi \!\circ\! \Bigg(
            \bigotimes_{i=1}^n \Gamma_i
        \Bigg)
        \Bigg(
            \bigotimes_{i=1}^n \rho \otimes \eta_i
        \Bigg)
    \Bigg] \, .
\end{align}
Notice that the measurements $E_k$  act separately and identically on each of the $n$ reservoirs.

The evolution of the system can be expressed in terms of a single CPTP map $\Omega$, which encapsulates the effects of both $\Gamma_i$ and $\Phi$ (i.e., $\Omega = \Phi \circ \left( \otimes_{i=1}^n \Gamma_i \right)$), so

\begin{equation}
    y_l = \sum_{k} W_{lk} \,
    \mathrm{Tr} \!\Bigg[
        E_k \,
        \Omega \!\left(
            \bigotimes_{i=1}^n \rho \otimes \eta_i
        \right)
    \Bigg] \, .
\end{equation}

Since all input states are indistinguishable, rather than explicitly associating each copy of $\rho$ with a specific reservoir, we can redefine the system on which the map acts as $\rho^{\otimes n} \otimes \eta_1 \otimes ... \otimes \eta_n$. This is possible due to the canonical tensor product isomorphism, which allows us to reorder factors in a tensor product space without altering the underlying structure of the quantum system. Thus, the output can now be rewritten in a more compact form by applying a different map, $\Tilde{\Omega}$, to this reordered system (i.e., $(\rho^{\otimes n} \bigotimes_{i=1}^{n} \eta_i)$).

In terms of effective measurements, 
\begin{equation}
\begin{split}
    y_l 
    &= \sum_{k} W_{kl} \, 
       \mathrm{Tr} \!\Big[
           E_k \, \Tilde{\Omega}\!\Big(
               \rho^{\otimes n} \bigotimes_{i=1}^{n} \eta_i
           \Big)
       \Big] \\[6pt]
    &= \sum_{k} W_{kl} \, 
       \mathrm{Tr}_S [
           \Tilde{E}_k \, \rho^{\otimes n}
       ] = \mathrm{Tr}_S [
           O_l \, \rho^{\otimes n}] ,
\end{split}
\end{equation}
where $\Tilde{E}_k = \mathrm{Tr}_{R} [ \Tilde{\Omega}^{\dagger} [E_{k}](\mathbb{I}_S^{\otimes n} \bigotimes_{i= 1}^n \eta_i)]$. It follows that with this architecture, it is also possible to reconstruct nonlinear functionals of the input state. Once again, the operator $O_l$ acts on a space of dimension $d_{s n}$, meaning that the required number of independent PVM outcomes is $p \geq d_{sn} - 1$. Taking this into account, and recalling that the measurements are performed on the entire entangled system, the required condition for the reservoir dimension is
\begin{equation}
\prod_{i=1}^n \dim(\mathcal{H}_{R_i})\dim(\mathcal{H}_S) -1 \geq d_{s n} - 1 .
\end{equation}

If we take all reservoirs to be of the same dimension, it becomes
\begin{equation}
\dim(\mathcal{H}_{R}) \geq \frac{\sqrt[n]{d_{sn}}}{\dim(\mathcal{H}_{S})} .
\label{eq:d}
\end{equation}

In comparison to Eq.~\eqref{eq:mi}, the use of multiple reservoirs introduces the $n$-th root, reducing the number of qubits required in each reservoir, as detailed in \ref{sec:scaling}. Furthermore, $n$ distributed (entangled) reservoirs allow addressing non-linear functions up to degree $n$.


\section{Scaling Analysis}
\label{sec:scaling}

\begin{figure}[t!]
    \centering

    \begin{subfigure}[t]{0.9\linewidth}
        \centering
        \includegraphics[width=\linewidth]{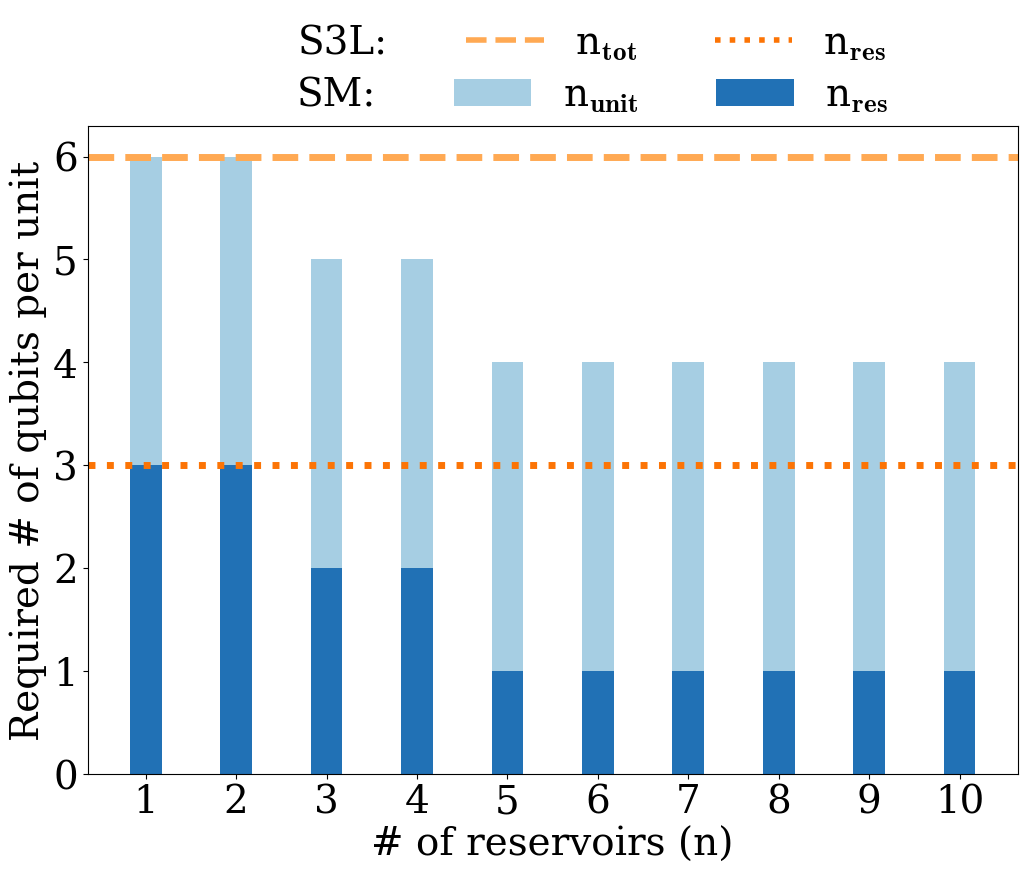}
        \put(-220,180){\textbf{(a)}}
        \label{fig:sc1}
    \end{subfigure}

    \vspace{0.4cm} 

    \begin{subfigure}[t]{0.9\linewidth}
        \centering
        \includegraphics[width=\linewidth]{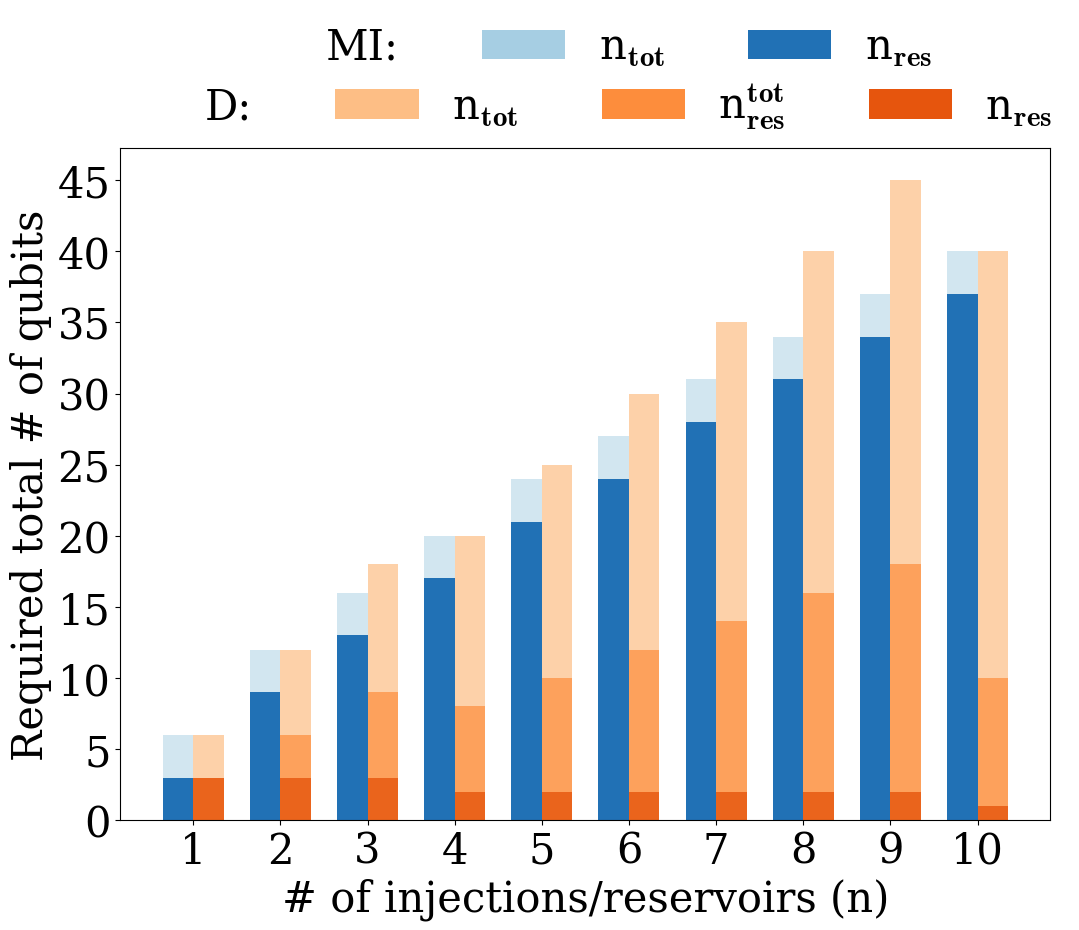}
        \put(-220,180){\textbf{(b)}}
        \label{fig:sc2}
    \end{subfigure}
    \caption{Qubit requirements across different architectures for a three-qubit input. (a) Comparison between the single three-layer (S3L) and spatially multiplexed (SM) architectures for linear property reconstruction. In S3L, $n_{res}$ denotes the number of qubits in the reservoir, and $n_{tot}$ the total number of qubits used. In SM, $n_{res}$ represents the number of qubits in a single reservoir and $n_{unit}$ the total number of qubits per unit, including the input. (b) Comparison of the multiple-injections (MI) and distributed (D) architectures for nonlinear property reconstruction. In MI, $n_{res}$ represents the number of qubits in the reservoir, while $n_{tot}$ denotes the total number of qubits, including the input. In D, $n_{res}$ represents the number of qubits in a single reservoir, $n_{res}^{tot}$ denotes the total number of qubits across all reservoirs, and $n_{tot}$ represents the overall number of qubits, including all considered inputs.}
    \label{fig:sc}
\end{figure}

In this section, we analyze the amount of resources required in the introduced architectures, comparing those with equivalent capabilities---namely, the ability to address linear and nonlinear tasks. For clarity, we use the required number of qubits rather than the Hilbert-space dimension, determined by the minimal qubit count whose associated dimension meets the bound. We focus on a three-qubit input; for generalization to other input sizes, see Appendix \ref{sec:app_colormaps}.

First, we compare the qubit requirements of the single three-layer architecture with those of the spatially multiplexed architecture, as both enable the reconstruction of linear properties. Figure \ref{fig:sc}(a) shows how, in spatial multiplexing, increasing the number of units reduces the number of qubits required per reservoir and, consequently, the total number of qubits per unit. Compared to the fixed number of qubits required in the single three-layer architecture, the difference becomes increasingly significant as the number of units grows. For instance, in this specific example, a single reservoir with three qubits would be equivalent to using 5 reservoirs with 1 qubit each. While the total number of processed qubits is smaller in the case of a single reservoir, the advantage of the spatially multiplexed QELM lies in the possibility of using limited resources in each (parallel) experiment. Indeed, each unit in this architecture operates independently, allowing implementation in separate experimental runs and avoiding the need to scale up a single quantum device. From the analytical expressions in Eqs. \ref{eq:single} and \ref{eq:sm}, this corresponds to a linear improvement in the required reservoir dimension, since for large $N_q$-qubit inputs it scales as $2^{N_q}/n$, compared to the $2^{N_q}$ for the single three-layer architecture. Consequently, spatial multiplexing offers not only resource efficiency but also experimental flexibility.

On the other hand, for nonlinear properties, in Figure \ref{fig:sc}(b) we compare the multiple-injections and distributed architectures. The presence of entanglement among units in the distributed architecture implies that the qubit requirement must be evaluated for the entire system rather than for individual units. Taking this into account, increasing the number of units leads to a higher qubit requirement, just as when increasing the number of injections in the multiple-injections architecture. This is because introducing the input state $n$ times means that we aim to reconstruct a higher-dimensional Hilbert space $\mathcal{H}_S^{\otimes n}$, in order to capture higher-order nonlinearities. In fact, since both architectures aim to cover the same space, they require nearly the same total number of qubits, despite their structural differences. Minor discrepancies —--where the distributed architecture may use slightly more qubits—-- stem from the imposed constraint that each reservoir has the same size, which sometimes forces overallocation of qubits. For example, when 9 reservoirs are used and the total reservoir requirement is 10 qubits, assigning 1 qubit to each reservoir would not be enough, while assigning 2 qubits to each would result in far above the minimum necessary. In contrast, when considering 10 reservoirs with the same 10-qubit requirement, allocating 1 qubit per reservoir perfectly matches the total, resulting in a lower overall qubit count than in the case of nine reservoirs. This effect can be seen in Figure\ref{fig:sc}(b), although such cases can be easily identified and corrected by simply removing 1 qubit from the appropriate reservoirs. The analysis presented here is intended to provide a general overview of the scaling behavior, which is still captured even under the uniform-reservoir constraint imposed for simplicity. 

Although they have the same overall requirements, in the multiple-injection design, the number of qubits increases because the dimension of the single reservoir grows, whereas in the distributed architecture, multiple small reservoirs are used, each dedicated to processing a separate input (as highlighted in Figure\ref{fig:sc}(b)). Therefore, a key advantage of the distributed approach is its division into independent entangled subsystems, each consisting of a small reservoir and its corresponding input, which can be measured separately. This makes the scheme experimentally practical, in contrast with the presence of a large reservoir in the multiple-injection case. Moreover, while the latter processes injections sequentially, the distributed scheme enables all subsystems to operate in parallel, offering potential gains in efficiency. The entanglement between reservoirs, potentially even in different labs, could be achieved for instance by entanglement routing over a quantum network \cite{pant_routing_2019}.

These considerations are not limited to the QELM framework but are also applicable when designing distributed protocols for quantum state reconstruction using other models that rely on enlarging the measurement space.

\section{Numerical Results}
\label{sec:results}

In this section, we present the numerical results demonstrating the performance of the proposed architectures in reconstructing a variety of quantum targets. For each experiment, a dataset of $N$ randomly generated density matrices and their associated targets is constructed, split into 80\% for training ($N_{\mathrm{tr}}$) and 20\% for testing ($N_{\mathrm{ts}}$). Note that, to guarantee numerical stability in the linear regression, the number of training samples must exceed the number of PVM outcomes, which is directly related to the dimension of the input space to be reconstructed.

Each input state is evolved together with a reservoir, initialized in a randomly chosen mixed state, under the all-to-all transverse Ising model 
\begin{equation}
    H = \frac{1}{2} \sum_{i > j} J_{ij} \sigma_x^{(i)}\sigma_x^{(j)} + h \sum_i \sigma_z^{(i)},
    \label{eq:ham}
\end{equation}
where $\sigma_\alpha^{(i)}$, $\alpha = x, y, z$ identifies the Pauli operator on the $i-th$ particle. The couplings $J_{ij}$ are randomly chosen from a uniform distribution in the interval $[-J_s, J_s]$, with $J_s = 1$, while $h$ is set to $h = 1$. Note that this Hamiltonian acts on the qubits belonging to both the input and reservoir systems. With this, it is possible to construct the evolution operator U=$e^{-iHt}$, which defines the CPTP map as $\Gamma = U (\cdot) U^\dagger$. This map is applied to the system $\rho \otimes \eta$, which is then measured with the set of PVMs $\{E_k\}$. We set the evolution time to $t=10$ because, together with our parameter choices, this places the system in the ergodic regime, where information spreads efficiently. Such dynamics are known to improve reservoir performance and still retain enough locally accessible information to enable accurate quantum state reconstruction \cite{rodrigo, vetrano_state_2024}. In architectures with multiple reservoirs, different maps correspond to different values of $J$ in the Hamiltonian. Furthermore, in the distributed architecture, an additional global map acting across all reservoirs is required to generate entanglement among them.

Each input state is evolved in conjunction with a reservoir, and careful consideration is given to the selected dynamical regime. A nonzero coupling strength ($J \neq 0$) is required to induce correlations and a local field ($h> 0$) is essential to produce quantum coherences, thereby ensuring that the reconstruction performance does not depend critically on the choice of measurement basis. In the present implementation, PVMs are consistently performed in the computational basis. Further details concerning the dynamical model are provided in Appendix \ref{sec:dynamics}.


In order to quantify the performance of the QELM, we use the following Normalized Mean Squared Error (NMSE) between the expected target $y_l$ and the prediction of the model $\hat{y}_l$ in the test step, where $\bm{y}_l$ now denotes the vector containing all samples corresponding to the $l$-th target.
\begin{equation}
    NMSE = \frac{\frac{1}{N} \sum_{i=1}^{N} ((y_i)_l-(\hat{y}_i)_l)^2}{Var(\bm{y}_l)}
\end{equation}


The results shown in this section are computed over 100 independent experimental runs using a dataset of size $N=200$ (unless stated otherwise).

\subsection{Required reservoir dimension}

We first address the ability of QELM in different designs to saturate the bounds discussed above. We analyze how the reconstruction error depends on the number of PVM outcomes and determine the point at which perfect reconstruction is achieved (i.e., NMSE within the numerical precision of the simulation).

\begin{figure}[h!]
    \centering
    \includegraphics[width=\linewidth]{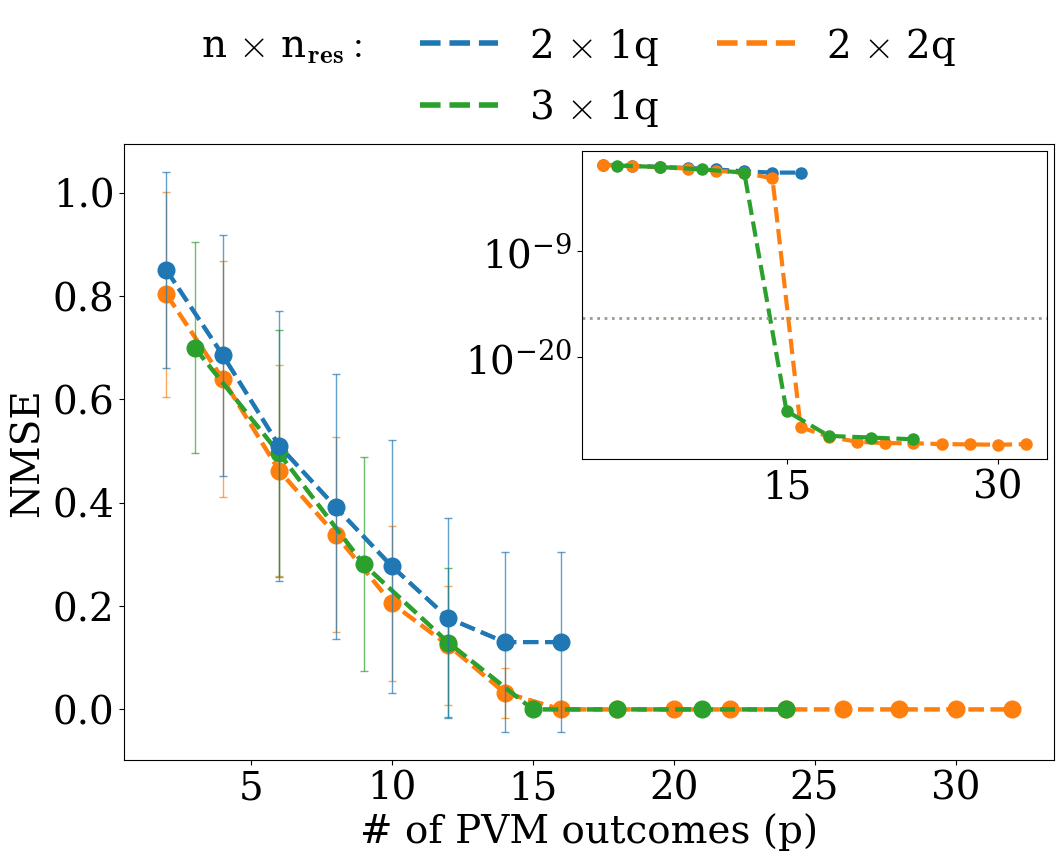}
    \caption{Reconstruction error in estimating $Tr[ (\sigma_x \otimes \sigma_x) \rho ]$ for a 2-qubit input using the spatially multiplexed architecture, evaluated for varying numbers of reservoirs ($n$) and qubits per reservoir ($n_{res}$). The error is depicted as a function of the number of available PVM outcomes obtained from measurements in the computational basis $\Pi_z$, which changes depending on the number of reservoirs and the number of qubits in each. The coupling parameter is sampled from $J \in [-1,1]$ and h=1. Each point represents the mean over 100 experimental runs, with bars indicating the standard deviation. The inset shows the mean values on a logarithmic scale, highlighting the point at which perfect reconstruction is achieved.}
    \label{fig:smex}
\end{figure}

To this end, we focus on the spatially multiplexed architecture and the reconstruction of linear targets of the form $y=\mathrm{Tr} [O \rho]$. Since the number of available outcomes depends on the number of reservoirs and their dimension, we attempt to reconstruct the same target using different configurations to confirm that the behavior is as expected. Specifically, we consider an example of a two-qubit input for which we want to reconstruct $\mathrm{Tr}[ (\sigma_x \otimes \sigma_x) \rho ]$. In this case, the input dimension requires at least 15 independent PVM outcomes, and the number of qubits needed in each reservoir depends on the number of reservoirs considered, as given in Eq.~\eqref{eq:sm}. In Figure \ref{fig:smex}, we present the reconstruction error for configurations with two reservoirs containing one and two qubits each, as well as for an alternative configuration with three reservoirs containing one qubit each. The results show that using two reservoirs with a single qubit each does not allow perfect reconstruction since, although 16 PVM outcomes are obtained, only 14 are independent (i.e., one outcome per unit is redundant). When an additional qubit is added to each reservoir, perfect reconstruction becomes achievable, in agreement with the theoretical bound. Adding a third reservoir also makes it possible to achieve accurate reconstruction with only a single qubit per reservoir, corresponding to another theoretically valid configuration.

This behavior is consistently observed across different observables and architectures. The reconstruction error vanishes as soon as the number of PVM outcomes reaches the minimum required, satisfied when the reservoir dimension meets the provided theoretical bounds. Additional results confirming this behavior for the other architectures are provided in Appendix \ref{sec:bounds}. Therefore, in the following, we set the number of qubits in each reservoir to the minimum required, which depends on the architecture used.

\subsection{Nonlinear targets}

\begin{figure}[h!]
    \centering
    \includegraphics[width=\linewidth]{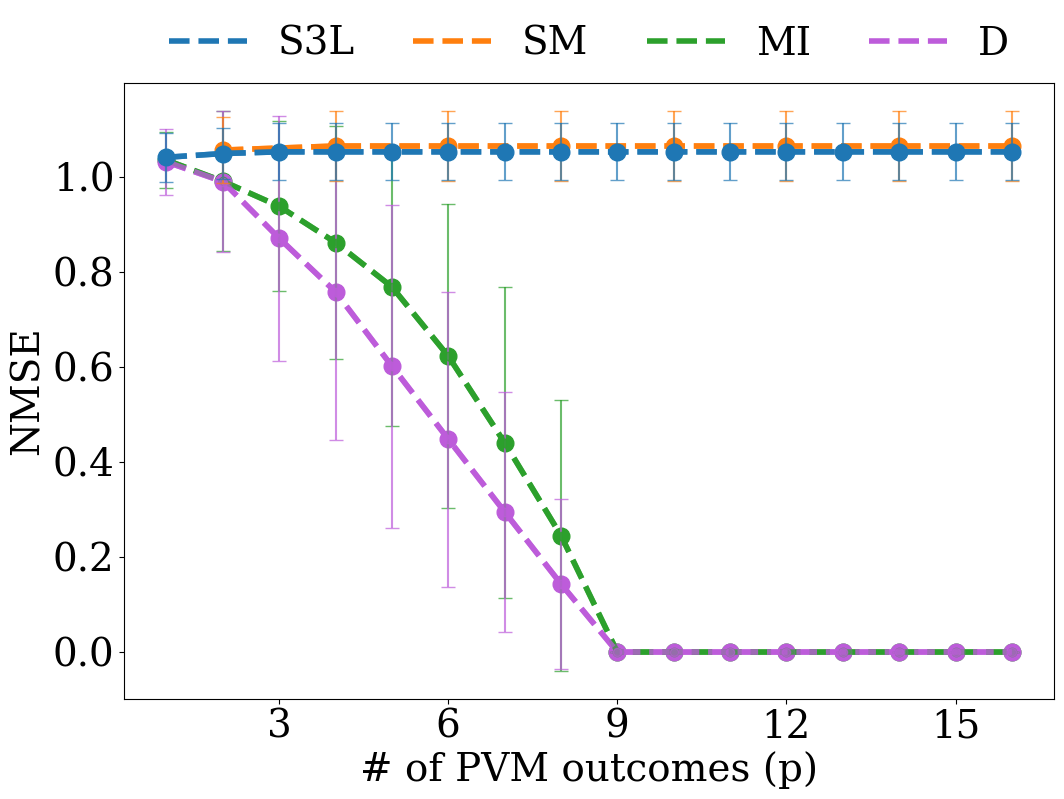}
    \caption{Reconstruction error in estimating the purity of a 1-qubit input using all the architectures: single three-layer (S3L), spatially multiplexed (SM), multiple-injections (MI), and distributed (D). The number of qubits in the reservoirs is selected so that all architectures yield the same number of PVM outcomes, with SM and D using 2 reservoirs each and MI performing 2 input injections to enable the reconstruction of second-order nonlinearities. The measurement basis, parameter choices, and statistical representation are the same as in Figure \ref{fig:smex}.} 
    \label{fig:ex3}
\end{figure}

Nonlinear targets are of particular interest because many relevant properties of quantum states are inherently nonlinear functions of the density matrix. However, accurately reconstructing them is considerably more challenging. From their design, the single three-layer and spatially multiplexed architectures are fundamentally limited in this task, while the multiple-injection and distributed architectures are expected to achieve nonlinearity up to a degree depending on the number of copies of the input state injected. This is confirmed by the results illustrated in Figure \ref{fig:ex3}, where the purity of the input state, $Tr[ \rho^2 ]$, is the target.

In the following, we focus on our proposed distributed architecture to investigate the extent to which it enables the reconstruction of a broader class of nonlinear properties, including polynomial targets, entropy, and entanglement. 

\subsubsection{Polynomial targets}
\label{sec:pol}

Polynomial targets are of the form $\mathrm{Tr}[O\rho^k]$, where $O$ is a fixed observable and $k$ determines the degree of nonlinearity with respect to the input state $\rho$. Such functions play a key role in quantum error mitigation techniques, such as virtual distillation \cite{Huggins}.

\begin{figure}[h!]
    \centering
    \includegraphics[width=\linewidth]{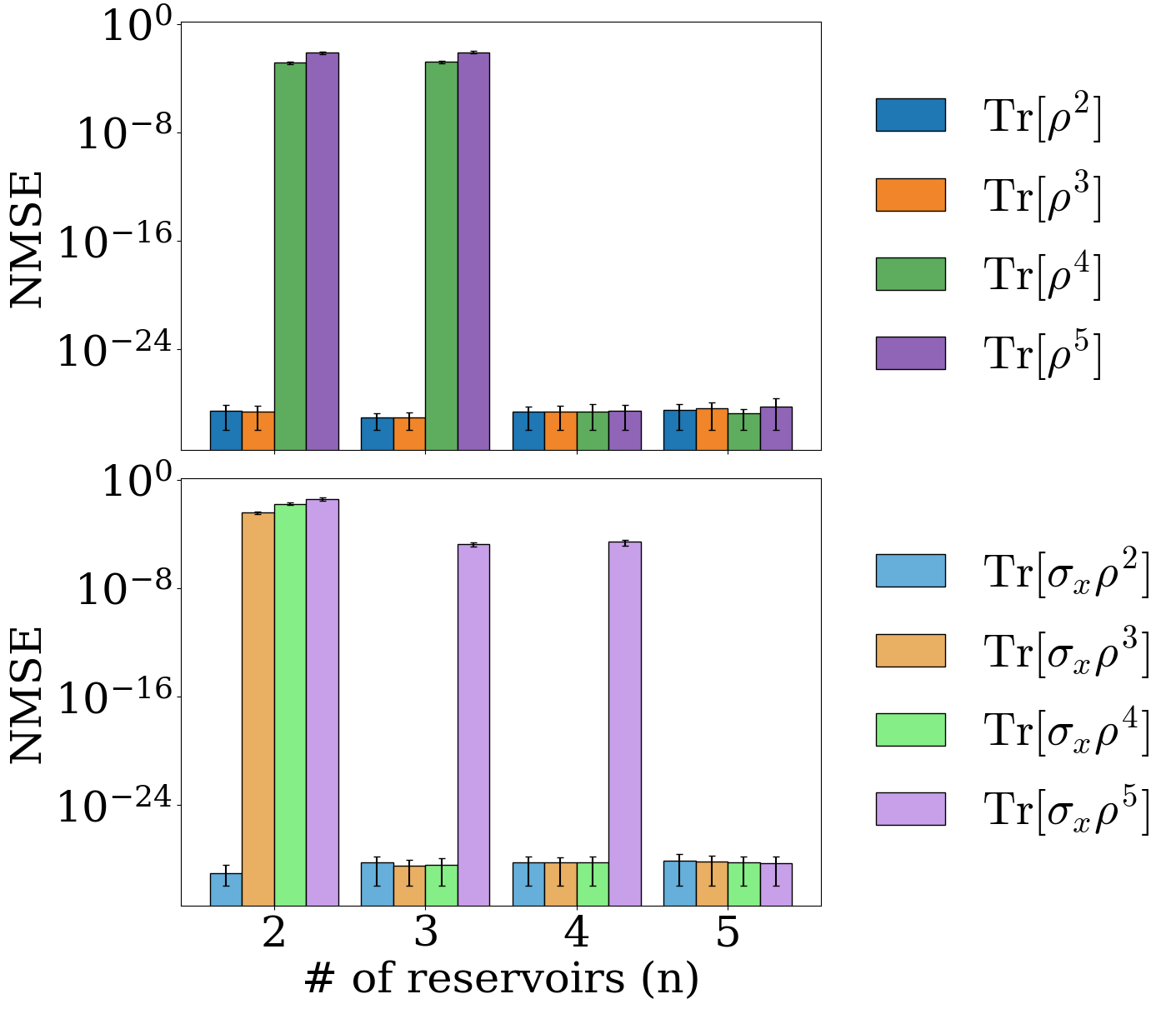}
    \caption{Reconstruction error in estimating polynomial targets of the form $Tr[O\rho^k]$ for a 1-qubit input, using the distributed architecture with a varying number of reservoirs. The number of qubits in each reservoir is adjusted according to the total number of reservoirs used. The top panel shows results for $O=I$ and the bottom panel for $O=\sigma_x$, with degree $k=2,3,4,5$ in each case. The measurement basis, parameter choices, and statistical representation are the same as in Figure \ref{fig:smex}.}
    \label{fig:full}
\end{figure}

While one might expect that $k$ reservoirs are always required to reconstruct a target of degree $k$, the actual number depends non-trivially also on the observable $O$. What we guarantee with $k$ reservoirs is the ability to access all the information contained in $\rho^k$, effectively enabling the reconstruction of any quantity of the form $\mathrm{Tr}[O\rho^k]$. However, certain targets may still be reconstructed with fewer reservoirs. This is what is represented in Figure \ref{fig:full}, where the nonlinearity introduced by the two-reservoir distributed architecture is sufficient to reconstruct $Tr[\rho^3]$, but not $Tr[\sigma_x\rho^3]$, which requires the introduction of a third reservoir. With three reservoirs, the situation reverses: $Tr[\sigma_x\rho^4]$ becomes accessible, while $Tr[\rho^4]$ does not. The general trend observed is that when $O=I$, reconstructing the targets with even $k$ requires $k$ reservoirs, whereas for odd $k$, $k-1$ reservoirs are sufficient. Conversely, when $O=\sigma_x$, targets with even $k$ can be reconstructed just with $k-1$ reservoirs, while odd $k$ requires $k$. Importantly, using 5 reservoirs allows the reconstruction of any target with $k \leq 5$. This occurs because the observable's actual degree of nonlinearity, which can be lower than the power $k$, determines the required resources, as detailed in Appendix \ref{sec:full}.

\subsubsection{Rényi entropy}

Entropy is a fundamental to characterize quantum states, but directly accessing it can be experimentally demanding. Specifically, the Rényi entropy is a generalized measure of quantum uncertainty given by
\begin{equation}
    S_{\alpha}(\rho) = \frac{1}{1-\alpha}log(Tr[\rho^{\alpha}]).
\end{equation}
where $\alpha > 0$ and $\alpha \neq 1$. When $\alpha \rightarrow 1$, it recovers the von Neumann entropy, which represents the conventional measure of mixedness for a quantum state.

\begin{figure}[h!]
    \centering
    \includegraphics[width=\linewidth]{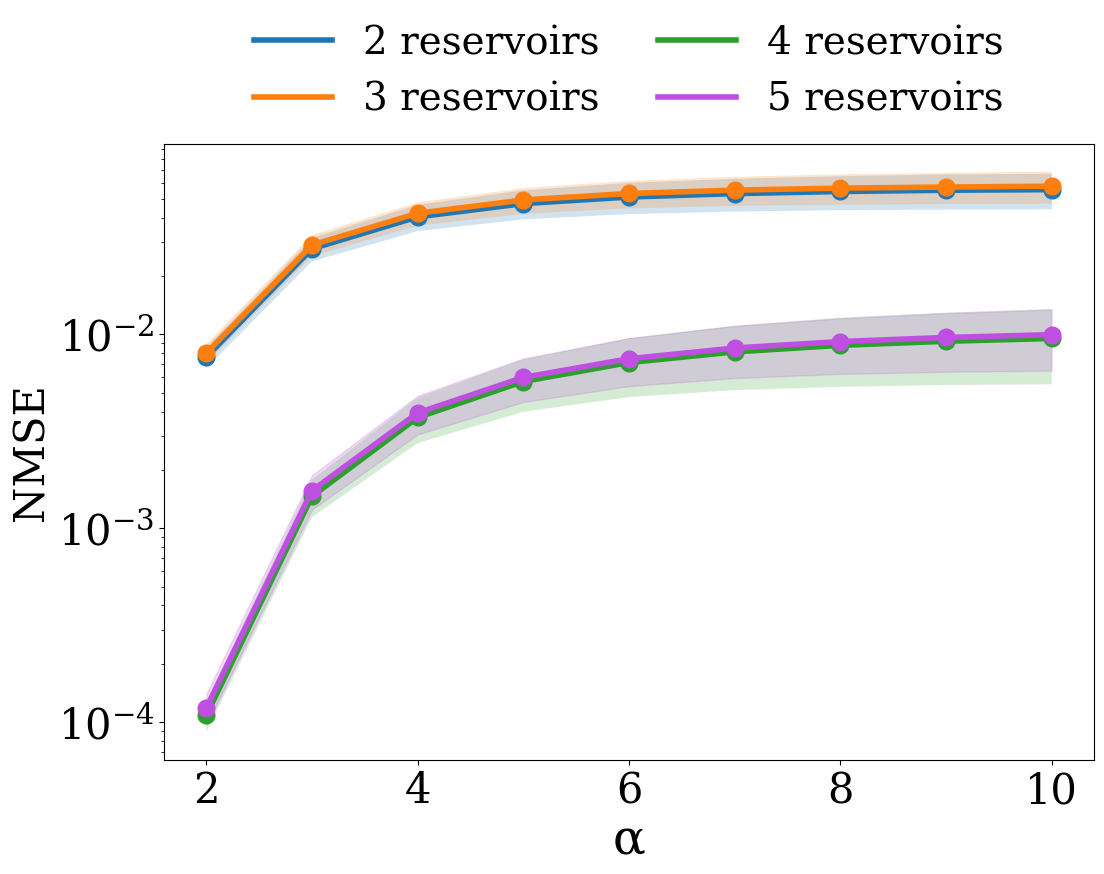}
    \caption{Reconstruction error in estimating the Rényi entropy $S_{\alpha}(\rho)$ for a 1-qubit input and various values of $\alpha$, using the distributed architecture with different numbers of reservoirs. The number of qubits in each reservoir is adjusted according to the total number of reservoirs used. The plotted values correspond to the mean NMSE over 100 independent experiments with a dataset of size $N=500$, with shaded regions indicating the corresponding standard deviations. The results for two and three reservoirs appear overlapped, as do those for 4 and 5 reservoirs. The measurement basis and parameter choices are the same as in Figure \ref{fig:smex}.}
    \label{fig:nonlinear1}
\end{figure}

Figure \ref{fig:nonlinear1} shows the reconstruction error obtained for a 1-qubit input when estimating the Rényi entropy as a function of the parameter $\alpha$, using up to five reservoirs. As $\alpha$ increases, the function becomes increasingly nonlinear, making the reconstruction task more challenging and resulting in larger errors. Even for $\alpha = 2$, perfect reconstruction cannot be achieved with this number of reservoirs due to the additional complexity introduced by the logarithmic transformation. Nevertheless, increasing the number of reservoirs consistently reduces the reconstruction error. 
Notably, only an even number of reservoirs provides improvement, as odd configurations yield errors equal to the preceding even ones. This effect arises because, as shown in Figure \ref{fig:full}, when $\alpha$ is odd, $\mathrm{Tr}[\rho^{\alpha}]$ exhibits the same reconstruction complexity as the preceding even $\alpha-1$, thereby introducing identical nonlinearities.

\subsubsection{Entanglement}

Finally, we aim to reconstruct entanglement measures, as entanglement is one of the most fundamental features of quantum systems and a key resource in quantum information processing. Although entanglement estimation simplifies to single-parameter inference for certain well-structured state families \cite{PhysRevA.111.022412}, its characterization for arbitrary quantum states is generally far more complex. Specifically, we focus on concurrence and negativity for 2-qubit systems, which are two of the most widely used measures of entanglement. 

The concurrence is defined as
\begin{equation}
    C(\rho) = \max (0, \lambda_1-\lambda_2-\lambda_3-\lambda_4),
    \label{eq:conc}
\end{equation}
where $\lambda_1,\lambda_2,\lambda_3,\lambda_4$ are the eigenvalues, in decreasing order, of the matrix $R=\sqrt{\sqrt{\rho}\tilde{\rho}\sqrt{\rho}}$, with $\tilde{\rho} = (\sigma_y \otimes \sigma_y)\rho^*(\sigma_y \otimes \sigma_y)$ and $\rho^*$ denoting the complex conjugate of $\rho$.

On the other hand, the negativity is given by
\begin{equation}
\mathcal{N}(\rho) = \frac{ \| \rho^{T_B} \|_1 - 1 }{2},
\end{equation}
where $\rho^{T_B}$ is the partial transpose of $\rho$ with respect to subsystem $B$ and $\|\cdot\|_1$ represents the trace norm.

Both are highly nonlinear functions, which makes their reconstruction particularly challenging, especially when using a small number of reservoirs \cite{zia_quantum_2025}. Here, we aim to address this challenge by employing the distributed architecture.

Table \ref{tab:ent-combined} shows the reconstruction error obtained when estimating both quantities for a 2-qubit input, using up to four reservoirs and an amount of training samples scaled to the dimension of the reconstructed space. Since each additional reservoir enlarges the space to be reconstructed, the required number of training samples must also increase in order to maintain numerical stability in the linear regression. For two reservoirs, the dimension is $d_{sn}=136$, implying a minimum of 136 training samples. Similarly, for three and four reservoirs, the corresponding dimensions are $d_{sn}=816$ and $d_{sn}=3876$, requiring at least that many samples, respectively. To ensure sufficient and comparable training conditions across these cases, we use 50\% more than the minimum number of required samples in each case. The results show that even though the errors are relatively large, there is a trend of decreasing error as the number of reservoirs increases, since higher orders of nonlinearity can be recovered. Notably, the standard deviations are larger in the cases with fewer reservoirs, suggesting that the small number of samples is insufficient to achieve a stable solution.

Another option is to fix the dataset size to be the same across all cases, setting it to 10000 samples to exceed the minimum required for the four-reservoir case. The reconstruction errors obtained under this configuration are also reported in Table \ref{tab:ent-combined}. The results indicate that the errors decrease with the number of samples and become more stable, as reflected by smaller standard deviations. The decreasing trend with increasing number of reservoirs is maintained, although the improvement becomes smaller, and no difference is found in the case of the negativity between three and four reservoirs.

\begin{table}[t]
  \centering
  \caption{Reconstruction error in estimating entanglement measures for a 2-qubit input, using the distributed architecture with different numbers of reservoirs. "$V$" is the case where the number of samples in the dataset is variable, chosen in proportion to the dimension of the space to be reconstructed, with $N=$255, 1530, and 7268 samples for two, three, and four reservoirs, respectively. "$F$" is the case where the number of samples is fixed to $N$=10000. The values correspond to the mean NMSE over 25 independent experiments, with the corresponding standard deviations.}
  \label{tab:ent-combined}

  \small
  \setlength{\tabcolsep}{6pt}

  \begin{tabular}{c |c |ccc}
    \toprule
    & \textbf{Tg.} & \multicolumn{3}{c}{\textbf{NMSE}} \\
    & & \textbf{2 reservoirs} & \textbf{3 reservoirs} & \textbf{4 reservoirs} \\
    \midrule

    \multirow{2}{*}{\textbf{V}}
      & $C(\rho)$ & $0.849\pm0.275$ & $0.370\pm0.046$ & $0.178\pm0.009$ \\
      & $\mathcal{N}(\rho)$  & $0.574\pm0.208$ & $0.179\pm0.022$ & $0.122\pm0.007$ \\
    \midrule

    \multirow{2}{*}{\textbf{F}}
      & $C(\rho)$ & $0.254\pm0.009$ & $0.123\pm0.003$ & $0.097\pm0.004$ \\
      & $\mathcal{N}(\rho)$  & $0.172\pm0.006$ & $0.060\pm0.002$ & $0.068\pm0.003$ \\
    \bottomrule
  \end{tabular}
\end{table}

A common finding is that the concurrence is more difficult to recover than negativity, resulting in larger reconstruction errors. Furthermore, the inclusion of a third reservoir has a greater impact than the addition of a fourth, particularly for the negativity. It is expected that adding more reservoirs would continue to decrease the error by enabling the recovery of higher-order nonlinearities, at the cost of a greater demand for computational resources.


\section{\label{sec:conc} Discussion and Conclusions}


The standard paradigm for the complete reconstruction of a quantum state is quantum state tomography; however, its measurement and computational costs scale unfavorably with system size, limiting its applicability to large quantum systems \cite{paris2004quantum}. In response to these challenges, alternative frameworks have been developed that aim to extract relevant information from quantum states without performing full tomography. Among these, shadow tomography enables the estimation of selected observables or properties using substantially fewer measurements \cite{aaronson}. A practical implementation of this idea is provided by classical shadows, which rely on randomized measurements to efficiently predict many features of a quantum state \cite{huang_predicting_2020}. Related strategies based on the use of ancilla-assisted measurements have also been investigated, particularly in the context of analog quantum simulators \cite{McGinley, tran}.
However, this method remains limited in practice, as it can efficiently estimate local and few-body observables, but becomes impractical for generic ones.
This is what makes the use of machine learning techniques, such as QELMs, particularly appealing, as they can learn any property directly from data. QELMs can outperform classical statistical methods when high-quality labeled training data are available, e.g., in experimental platforms capable of efficiently generating and labeling simple quantum states. Nevertheless, special attention must be paid to model architecture, as single-injection protocols generally allow only linear features to be extracted, whereas multiple-injection or distributed design schemes enable access to nonlinear properties and more complex tasks\cite{innocenti_potential_2023, peng2025, huang2022}.


In this work, we have studied how a quantum machine learning technique, such as a QELM, is capable of estimating arbitrary properties of quantum states by exploiting the enlarged Hilbert space generated by the use of the reservoir. Furthermore, motivated by the limitations of current quantum hardware, we have analyzed it in distributed contexts to assess how the required resources can be partitioned into smaller, more manageable subsystems. By explicitly linking architectural design choices to the class of accessible quantum properties, we identify the conditions under which different QELM architectures enable the reconstruction of observables, entropic quantities, and entanglement measures.

Compared with methods aimed at full state reconstruction, such as conventional tomography or informationally complete measurements (e.g., SIC-POVMs, mutually unbiased bases), the QELM approach offers a significantly simpler measurement strategy, relying only on PVMs in the computational basis. At the same time, it remains focused on predicting specific properties of the state, in the spirit of shadow tomography approaches. Treating the task as a machine learning problem provides substantial flexibility as the method does not impose structural restrictions on the target to be estimated, allowing any observable, local or global, to be learned using the same strategy. A further advantage is that, although the training phase may be computationally expensive, once the model is trained, it can be applied to any new state, as in any supervised learning protocol. 

To investigate these capabilities in detail, we have compared four QELM architectures, providing insights into the minimum reservoir sizes required to reconstruct arbitrary observables. Among the approaches considered, the single three-layer and the spatially multiplexed architectures are well suited for reconstructing linear functionals of the input, with the latter offering the advantage that increasing the number of units reduces the qubits required per reservoir, distributing the workload across smaller independent systems. Multiple injections of the input allow the reconstruction of increasingly complex targets, demonstrated by the multiple-injection architecture. Building on this idea, and motivated by the goal of overcoming hardware limitations through distributed schemes, we propose a distributed architecture that introduces entanglement between the units of the spatially multiplexed architecture. This architecture has been shown to require the same total number of qubits as the multiple-injections architecture, but in practice, it is composed of smaller entangled reservoirs that can be generated and measured separately and does not require the longer coherence times that can hinder multiple injections.

After scaling considerations, our proposed architecture has been applied to the reconstruction of several nonlinear targets to evaluate its capacity. For polynomial targets, it has been shown that any target exhibiting a nonlinearity of order $k$ can be reconstructed using $k$ reservoirs, although specific targets can be recovered with fewer reservoirs. When considering more complex targets, such as the Rényi entropy or entanglement measures, perfect reconstruction is not achieved with the limited number of reservoirs used. However, the error consistently decreases as the number of reservoirs increases, as expected from the recovery of higher-order nonlinearities.

As future work,
 we propose a systematic comparison of QELM-based reconstruction with alternative measurement strategies, such as SIC-POVM-based tomography or shadow tomography. It would be particularly valuable to assess the trade-offs between these methods in terms of measurement complexity, classical post-processing, and scalability, especially when they are embedded in spatially multiplexed or distributed architectures. This would clarify the regimes in which direct learning of target quantities, rather than full-state reconstruction followed by post-processing, offers a genuine practical advantage. In addition, understanding the noise robustness of these models is crucial for determining whether direct estimation of the target via more expressive architectures is advantageous compared to full-state reconstruction and subsequent post-processing. Beyond numerical simulations, the appeal of this approach is in its testing in experiments in NISQ quantum hardware platforms, such as photonic, multi-qubit superconducting or trapped-ion setups, capable of generating the required intra- and inter-reservoir entangling operations,  which would allow us to evaluate the method under realistic noise profiles and hardware-specific constraints. Finally, all the ideas of spatial multiplexing, repeated injection, and entangled subsystems are not limited to static states. A promising generalization is their adaptation to Quantum Reservoir Computing (QRC) for processing temporal quantum data streams, such as predicting the evolution of observables or quantum channel tomography.

 \vspace{1cm}

\begin{acknowledgments}

This work has been supported by the Gipuzkoa Provincial Council through the research grant KMAO, grant number 2025-QUAN-000027-01 (Algoritmo Kuantiko Berrien Garapena: Materialen Zientzia, Adimen Artifiziala eta Optimizazio Industriala). Furthermore, we acknowledge the Spanish State Research Agency, through the María de Maeztu project CEX2021-001164-M and through the COQUSY project PID2022-140506NB-C21 and -C22, all funded by MCIU/AEI/10.13039/501100011033; the project is funded under the Quantera II program that has received funding from the EU’s H2020 research and innovation program under the Grant No. 101017733 (QNET), and from the Spanish State Research Agency, PCI2024-153410 funded by MCIU/AEI/10.13039/50110001103; MINECO through the QUANTUM SPAIN project, and EU through the RTRP - NextGenerationEU within the framework of the Digital Spain 2025 Agenda. EF acknowledges support by the European Union’s Horizon Europe program HORIZON-MSCA-2022-PF-01, grant No. 101105267 (CoQHoNet). The CSIC Interdisciplinary Thematic Platform (PTI+) on Quantum Technologies in Spain (QTEP+) is also acknowledged. 

\end{acknowledgments}


\bibliography{Bib}
\bibliographystyle{unsrt}

\appendix

\section{Reconstruction under different dynamical regimes}
\label{sec:dynamics}

The ability to reconstruct certain targets depends on the dynamical regime induced by the system's Hamiltonian (Eq.~\eqref{eq:ham}).\@ Having both $J$ and $h$ nonzero is a necessary condition to reconstruct arbitrary observables from measurements in any basis, assuming that the dimensional bounds are fulfilled. 

\begin{figure}
    \centering
    \includegraphics[width=0.9\linewidth]{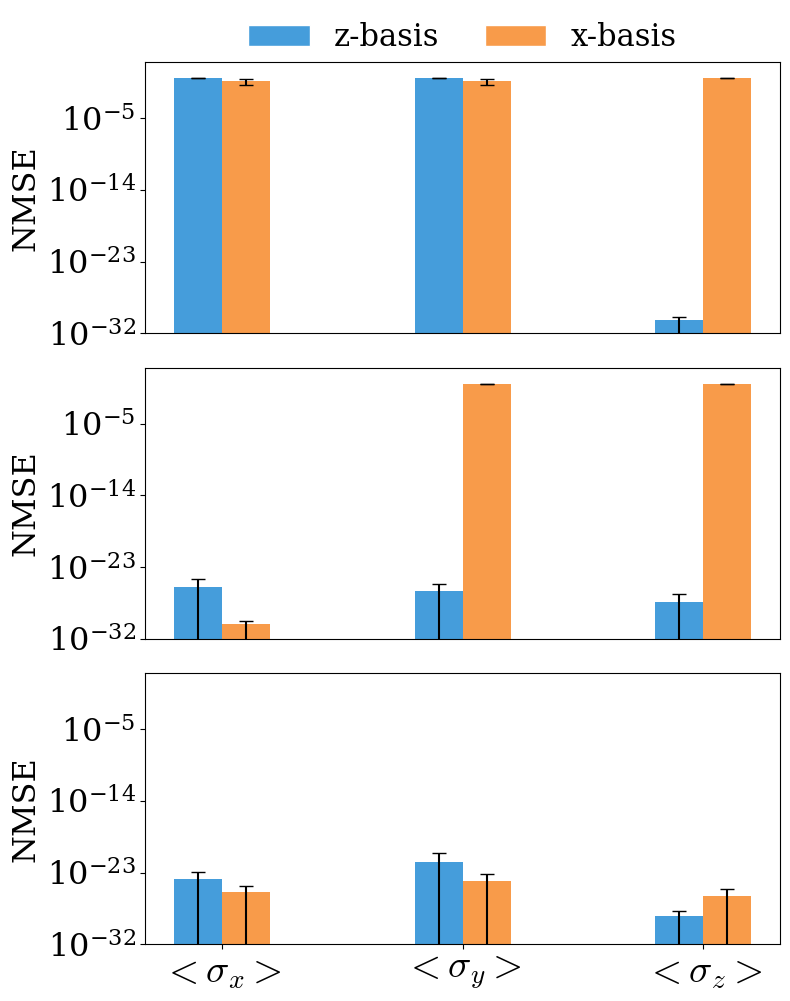}
    \caption{Reconstruction error in estimating the Pauli expectation values using the single three-layer architecture for a 1-qubit input under different dynamical regimes, with PVM measurements performed in either the computational or $x$-basis. Each point represents the mean over 100 experimental runs, with bars indicating the standard deviation. Top: $J=0$, $h \neq 0$. Middle: $J \neq 0$, $h=0$. Bottom: $J,h \neq 0$. When nonzero, the value of $J$ is chosen from $J \in [-1,1]$ and $h=1$.}
    \label{fig:dy}
\end{figure}

Clearly, when $J=0$, the reservoir is completely decoupled from the input, and performing the PVM in a single basis provides not enough outcomes, as in standard quantum state tomography.\@ Figure \ref{fig:dy} (top) illustrates this by showing the reconstruction error for $\mathrm{Tr}[\sigma_x\rho], \mathrm{Tr}[\sigma_y\rho]$, and $\mathrm{Tr}[\sigma_z\rho]$ for a 1-qubit input and a 1-qubit reservoir in the dynamical regime $J=0$, with measurements performed either in the computational (i.e., $\{\ket{00},\ket{01},\ket{10},\ket{11}\}$) or in the $x$-basis (i.e., $\{\ket{++},\ket{+-},\ket{-+},\ket{--}\}$). As expected, these observables can not be recovered from so few outcomes, except for $\mathrm{Tr}[\sigma_z\rho]$ when measuring in the computational basis because, in this setting, the dynamics reduces to a rotation about the $z$-axis, which leaves the input populations unchanged, and consequently the measurement outcomes directly provide these populations, whose difference yields the desired expectation value.

On the other hand, in Figure \ref{fig:dy} (middle) we consider $h=0$. In this case, the dynamics is aligned with a fixed interaction axis, limiting the visibility of input information in certain measurement bases. Adding local fields (i.e., $h \neq 0$) enriches the dynamics, allowing input features to be accessed regardless of the measurement direction. This can be understood by considering again a 1-qubit input $\rho = \frac{1}{2} (\mathbb{I}+\vec{r}_{\rho} \cdot \vec{\sigma})$ and a 1-qubit reservoir $\eta = \frac{1}{2} (\mathbb{I}+\vec{r}_{\eta} \cdot \vec{\sigma})$, where $\vec{\sigma}=(\sigma_x,\sigma_y,\sigma_z)$ is the vector of Pauli operators, and $\vec{r}_{\rho}=(r_x^{\rho}, r_y^{\rho}, r_z^{\rho})$, $\vec{r}_{\eta}=(r_x^{\eta}, r_y^{\eta}, r_z^{\eta})$ are the Bloch vectors of the respective states. 

Each component of a Bloch vector corresponds to the expectation value of the associated Pauli operator (i.e., $r_i=\mathrm{Tr}[\sigma_i\rho], i \in \{x,y,z\}$). Therefore, identifying the reconstructible observables reduces to analyzing the measurement outcomes after the system evolves, and determining which components of $\vec{r}_{\rho}$ appear in them. To proceed, we first write the joint initial state of the system as

\begin{equation}
    \rho \otimes \eta = \frac{1}{4} (\mathbb{I} \otimes \mathbb{I} + \sum_i r_i^{\rho} \cdot \sigma_i \otimes \mathbb{I} + \sum_j r_j^{\eta} \cdot \mathbb{I} \otimes \sigma_j + \sum_{i,j} r_i^{\rho} r_j^{\eta} \cdot \sigma_i \otimes \sigma_j),
\end{equation}
where $i,j \in \{x,y,z\}$. In the dynamical regime $h=0$, the time evolution is governed by the unitary operator

\begin{equation}
    U = e^{-iJt\sigma_x \otimes \sigma_x} = \cos(Jt)\mathbb{I} \otimes \mathbb{I}-i\sin(Jt)\sigma_x \otimes \sigma_x.
\end{equation}

After applying this unitary, the system evolves to $\rho(t)=U(\rho \otimes \eta)U^{\dagger}$, which is then measured in either the computational basis or the $x$-basis. Table \ref{tab:meas_vs_params} summarizes the outcomes. When measuring in the computational basis, all the expectation values $\mathrm{Tr}[\sigma_x\rho], \mathrm{Tr}[\sigma_y\rho]$, $\mathrm{Tr}[\sigma_z\rho]$ are accessible. On the other hand, measurements in the $x$-basis provide access only to $r_x^{\rho}$. This is consistent with Figure \ref{fig:dy} (middle), which shows the reconstruction error for each observable in this regime, highlighting the crucial role of the measurement basis.

\begin{table}[h!]
  \centering
  \caption{Presence of the input Bloch vector components $r_x^{\rho}, r_y^{\rho}, r_z^{\rho}$ in the PVM outcomes obtained from measurements in the computational and $x$-basis for the dynamical regime $h=0$. A $\checkmark$ indicates presence, while a $\times$ indicates absence.}
  \label{tab:meas_vs_params}
  \begin{tabular}{lcccc|cccc}
    \toprule
     & $P_{00}$ & $P_{01}$ & $P_{10}$ & $P_{11}$ & $P_{++}$ & $P_{+-}$ & $P_{-+}$ & $P_{--}$ \\
    \midrule
    $r_x^{\rho}$ & $\checkmark$ & $\checkmark$ & $\checkmark$ & $\checkmark$ & $\checkmark$ & $\checkmark$ & $\checkmark$ & $\checkmark$ \\
    $r_y^{\rho}$ & $\checkmark$ & $\checkmark$ & $\checkmark$ & $\checkmark$ & $\times$     & $\times$     & $\times$     & $\times$ \\
    $r_z^{\rho}$ & $\checkmark$ & $\checkmark$ & $\checkmark$ & $\checkmark$ & $\times$     & $\times$     & $\times$     & $\times$ \\
    \bottomrule
  \end{tabular}
\end{table}

To overcome these issues, we always consider the regime $J, h \neq 0$. In this case, the transfer of information to the reservoir ensures that enough PVM outcomes are generated to reconstruct any observable, and the result no longer depends on the choice of measurement basis, as shown in Figure \ref{fig:dy} (bottom).

For small but nonzero values of $J$, the dynamics does not produce entanglement but only classical correlations. This can be verified by examining the quantum mutual information, which quantifies the total amount of correlations (classical and quantum) between the two subsystems. It is given by

\begin{equation}
    I(A:B)=S(\rho_A)+S(\rho_B)-S(\rho_{AB}),
\end{equation}

where $S(\rho)=-\mathrm{Tr}(\rho log\rho)$ is the von Neumann entropy and $\rho_A=\mathrm{Tr}_B(\rho_{AB})$, $\rho_B=\mathrm{Tr}_A(\rho_{AB})$. If the mutual information is nonzero while the concurrence (Eq.~\ref{eq:conc}) remains zero, the correlations present in the state are purely classical. To examine whether such classical correlations are sufficient to reconstruct properties of the input, we attempt to recover a specific observable, $\mathrm{Tr}(\sigma_x\rho)$, for different values of $J$ (with $h \neq 0$). The results, shown in Table \ref{tab:corr}, reveal that for values of $J$ close to zero, the concurrence remains zero while the mutual information, though small, is nonzero. Even this weak amount of classical correlation is enough to achieve a low NMSE. Nevertheless, using the range $J \in [-1,1]$, entanglement is present in most cases.

\begin{table}[h!]
  \centering
  \caption{Reconstruction error in estimating $\mathrm{Tr}(\sigma_x\rho)$ for a 1-qubit input using the single three-layer architecture, evaluated for different values of $J$. Here, $|J|$ denotes the mean absolute value of the coupling, and $I(A:B)$ and $C(\rho)$ the mean mutual information and concurrence generated across the dataset.}
  \label{tab:corr}
  \begin{tabular}{c|cccc}
    \toprule
    \textbf{Target} & \textbf{$|J|$} & \textbf{$I(A:B)$} & \textbf{$C(\rho)$} & \textbf{NMSE}\\
    \midrule 
    \multirow{5}{*}{$<\sigma_x>$} & 0 & 0 & 0 & 1.04 \\
    & $2.60 \times 10^{-8}$ & $2.59 \times 10^{-14}$ & 0 & $2.27 \times 10^{-10}$ \\
    & $2.39 \times 10^{-6}$ & $3.28 \times 10^{-10}$ & 0 & $8.24 \times 10^{-15}$ \\
    & $2.59 \times 10^{-4}$ & $2.60 \times 10^{-6}$ & $3.64 \times 10^{-8}$ & $3.99 \times 10^{-17}$ \\
    & $0.024$ & $0.020$ & $0.019$ & $1.69 \times 10^{-22}$ \\
    \bottomrule
  \end{tabular}
\end{table}

\section{Qubit requirements for different input sizes}
\label{sec:app_colormaps}

\begin{figure}[h!]
    \centering
    \includegraphics[width=\linewidth]{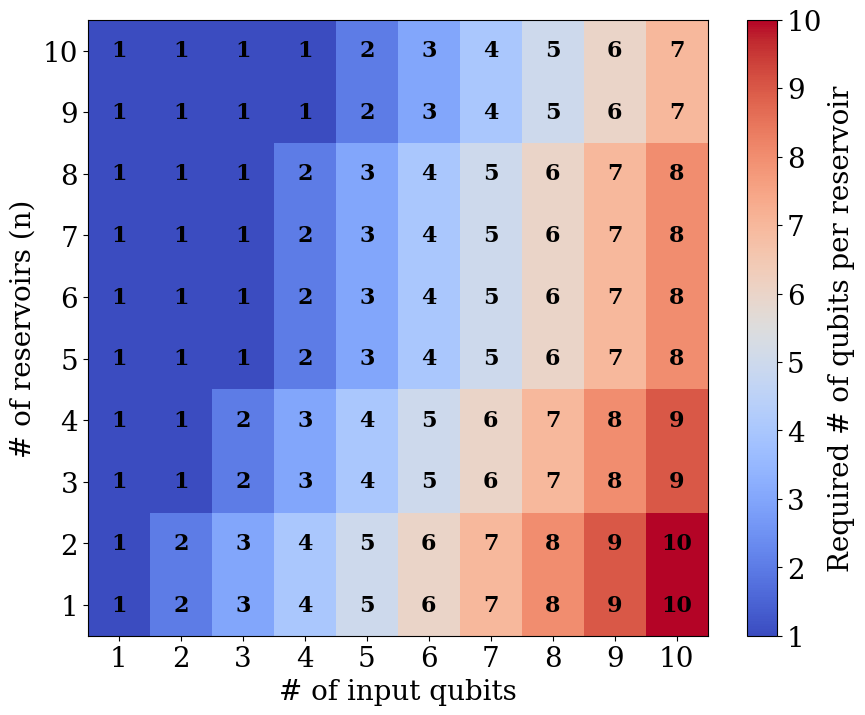}
    \caption{Required number of qubits per reservoir for different input sizes and varying numbers of reservoirs within the structure of the spatially multiplexed (SM) architecture.}
    \label{fig:colormap1}
\end{figure}

The following figures provide a more general overview of the qubit requirements for the architectures, now illustrating the results for input sizes ranging from 1 to 10 qubits and focusing on the number of qubits per reservoir. Figure \ref{fig:colormap1} corresponds to the spatially multiplexed architecture, where the required number of qubits per reservoir decreases as the number of reservoirs increases. 

\begin{figure}[h!]
    \centering
    \includegraphics[width=\linewidth]{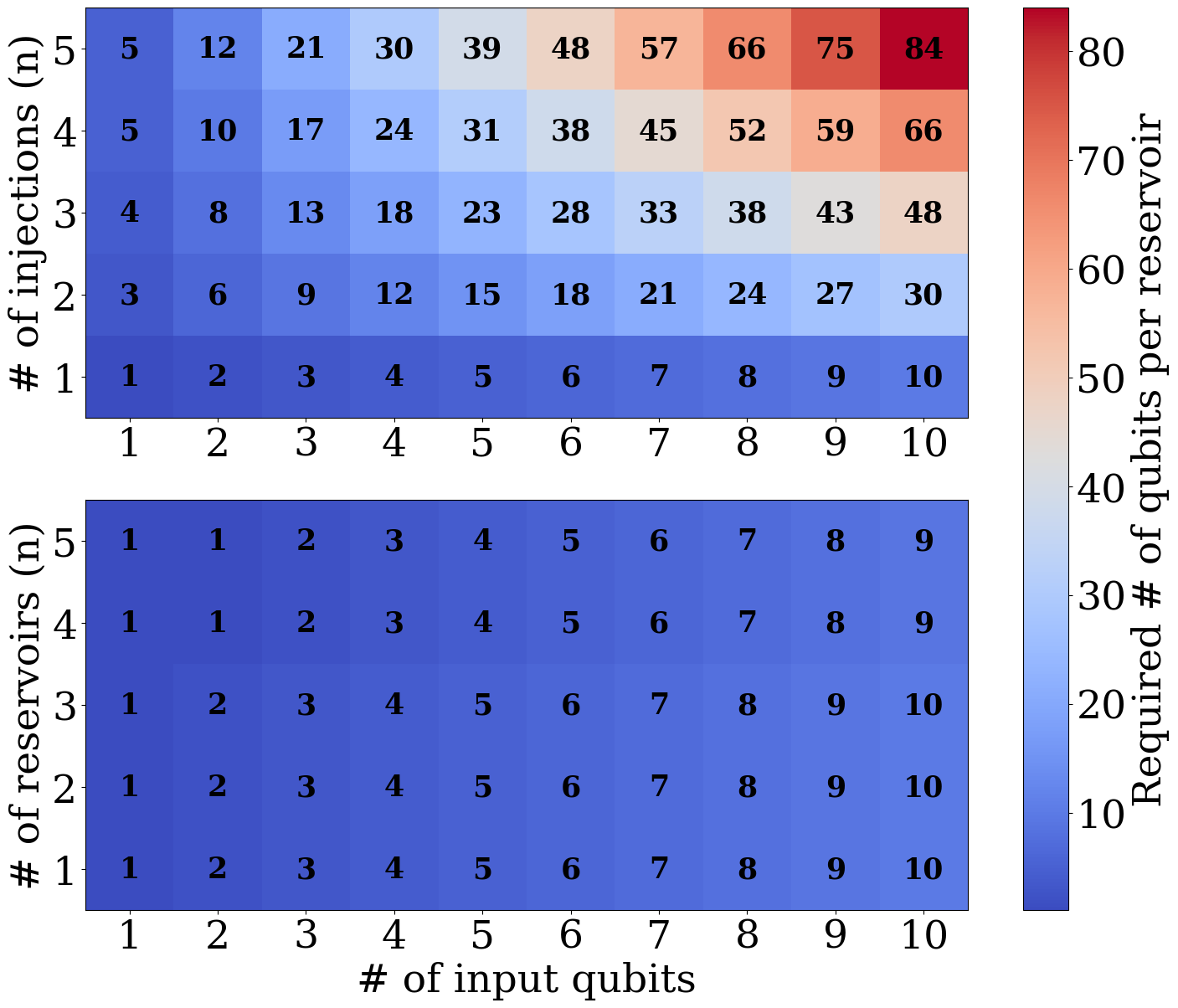}
    \caption{Top: Required number of qubits in the reservoir for different input sizes and varying number of injections in the multiple-injections (MI) architecture. Bottom: Required number of qubits in each reservoir for different input sizes and varying number of reservoirs in the distributed (D) architecture.}
    \label{fig:colormap2}
\end{figure}

On the other hand, in the multiple-injections architecture (Figure \ref{fig:colormap2} (top)), increasing the number of injections requires enlarging the reservoir to be able to cover all the space $\mathcal{H}_S^{\otimes n}$, which quickly leads to a large number of qubits for high-dimensional inputs. This contrasts with the distributed architecture (Figure \ref{fig:colormap2} (bottom)), keeping the qubit count in independent entangled subsystems small. Recall that this is possible because all input qubits are measured.

\section{Experimental validation of the theoretical bounds for different architectures}
\label{sec:bounds}

To illustrate the validity of the proposed theoretical bounds, we present several examples demonstrating that the architectures can successfully reconstruct the targets when these conditions are satisfied.

\begin{figure}[h!]
    \centering
    \includegraphics[width=\linewidth]{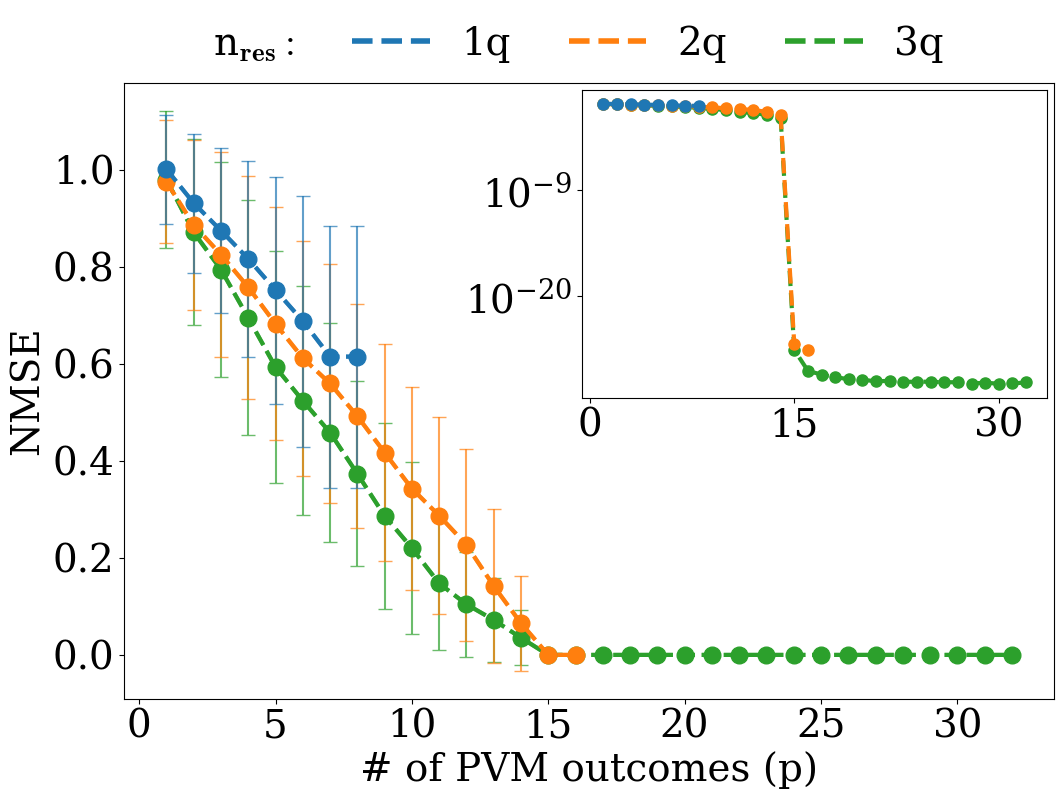}
    \caption{Reconstruction error in estimating $Tr[ (\sigma_x \otimes \sigma_x) \rho ]$ for a 2-qubit input using the single three-layer architecture, evaluated considering different numbers of qubits in the reservoir ($n_{res}$). The measurement basis, parameter choices, and statistical representation are the same as in Figure \ref{fig:smex}.}
    \label{fig:ex1}
\end{figure}

For the single three-layer architecture, we propose reconstructing the same target as in Figure \ref{fig:ex1} (i.e., $Tr[ (\sigma_x \otimes \sigma_x) \rho ]$ for a 2-qubit input). By varying the reservoir size, we observe that the results are consistent with the theoretical bound requiring the reservoir dimension to be at least equal to that of the input (Eq.~\eqref{eq:single}). With a one-qubit reservoir, the number of available PVM outcomes is insufficient to accurately estimate the target observable. In contrast, when using two or three qubits in the reservoir, the reconstruction error reaches zero, with this drop aligning with the expected requirement of 15 outcomes for a 2-qubit input. The results also suggest that increasing the reservoir size beyond the minimum required offers no apparent advantage, as perfect reconstruction is already achieved with the minimum number of qubits.

\begin{figure}[h!]
    \centering
    \includegraphics[width=\linewidth]{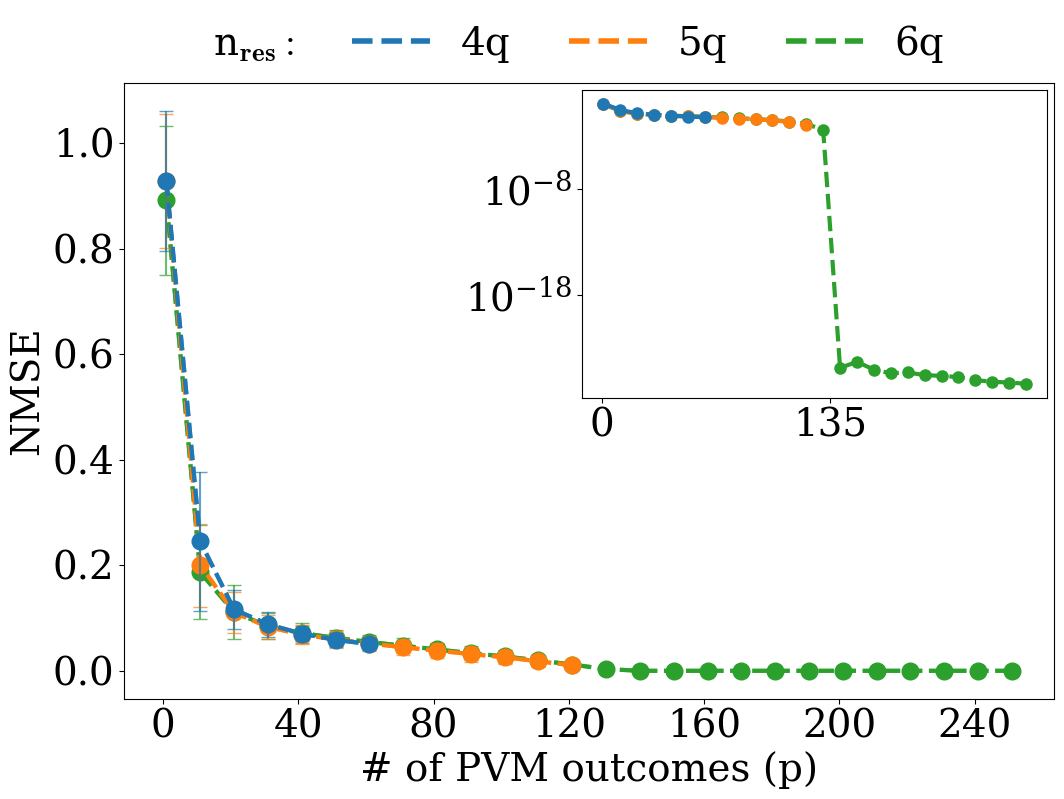}
    \caption{Reconstruction error in estimating $Tr[ (\sigma_x \otimes \sigma_x) \rho^2 ]$ for a 2-qubit input using the multiple-injections architecture, evaluated for a varying number of qubits per reservoir ($n_{res}$). The measurement basis, parameter choices, and statistical representation are the same as in Figure \ref{fig:smex}.}
    \label{fig:miex}
\end{figure}

\begin{figure}[h!]
    \centering
    \includegraphics[width=\linewidth]{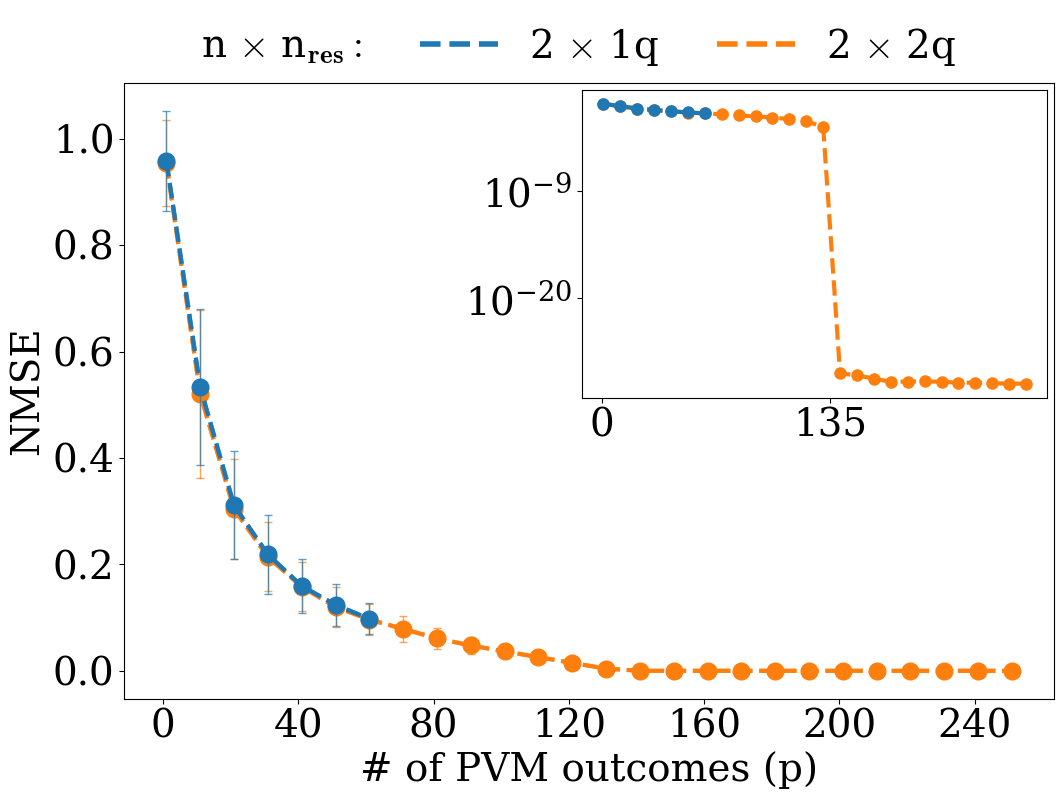}
    \caption{Reconstruction error in estimating $Tr[ (\sigma_x \otimes \sigma_x) \rho^2 ]$ for a 2-qubit input using the distributed architecture, evaluated for varying numbers of reservoirs ($n$) and qubits per reservoir ($n_{res}$). The measurement basis, parameter choices, and statistical representation are the same as in Figure \ref{fig:smex}.}
    \label{fig:dex}
\end{figure}

On the other hand, for the multiple-injections and distributed architectures, we consider reconstructing the target $Tr[ (\sigma_x \otimes \sigma_x) \rho^2 ]$ for a 2-qubit input. Since it involves a quadratic nonlinearity, we introduce the input twice to ensure that the entire space can be recovered. Taking this into account, $d_{sn}=136$, which implies that 135 independent PVM outcomes are required. In the multiple-injections architecture, as seen in Eq.~\eqref{eq:mi}, this corresponds to a 6-qubit reservoir, supported by Figure \ref{fig:miex}. When considering the distributed architecture, this corresponds to using 2 qubits in each reservoir, as given in Eq.~\eqref{eq:d} and confirmed in Figure \ref{fig:dex}.

\section{Distinguishing full and observable-specific reconstruction}
\label{sec:full}

As discussed in Section \ref{sec:pol}, when reconstructing $Tr[O\rho^k]$, the distributed architecture achieves zero error for any observable $O$ provided that the number of reservoirs equals $k$. However, even with fewer than $k$ reservoirs, it is sometimes possible to reconstruct specific targets. This can be understood by considering the general form of a single-qubit density matrix:

\begin{equation}
    \rho = \begin{bmatrix}
            p & q \\
            q^* & 1-p
            \end{bmatrix}
\end{equation}

As a specific example, using this form, we can compute $Tr[\rho^3]$ and $Tr[\rho^4]$. The expression for $\mathrm{Tr}[\rho^3]$ contains only quadratic terms, enabling reconstruction with just 2 reservoirs. In contrast, $ \mathrm{Tr}[\rho^4]$ includes terms up to fourth order, requiring 4 reservoirs.

\begin{equation}
    \mathrm{Tr}[\rho^3] = 3p^2-3p+3|q|^2+1
\end{equation}

\begin{align}
    \mathrm{Tr}[\rho^4] &= 2p^4 - 4p^3 + 4p^2|q|^2 + 6p^2 - 4p|q|^2 \notag \\
                          &\quad - 4p + 2|q|^4 + 4|q|^2 + 1
\end{align}

The situation is reversed when reconstructing the corresponding Pauli-X observables (i.e., $\mathrm{Tr}[\sigma_x\rho^3]$ and $\mathrm{Tr}[\sigma_x\rho^4]$). Both expressions now contain cubic terms, which require three reservoirs for reconstruction.

\begin{equation}
    \mathrm{Tr}[\sigma_x\rho^3] = (q+q^*)(p^2-p+|q|^2+1)
\end{equation}

\begin{equation}
    \mathrm{Tr}[\sigma_x\rho^4] = (q+q^*)(2p^2-2p+2|q|^2+1)
\end{equation}

This pattern holds for other even and odd values of $k$. For even $k$, reconstructing $\mathrm{Tr}[\rho^k]$ requires $k$ reservoirs, while $\mathrm{Tr}[\sigma_x\rho^k]$ can be reconstructed with $k-1$. Conversely, for odd $k$, $\mathrm{Tr}[\rho^k]$ requires only $k-1$ reservoirs, whereas reconstructing $\mathrm{Tr}[\sigma_x\rho^k]$ requires $k$.

\begin{figure}[h!]
    \centering
    \includegraphics[width=0.9\linewidth]{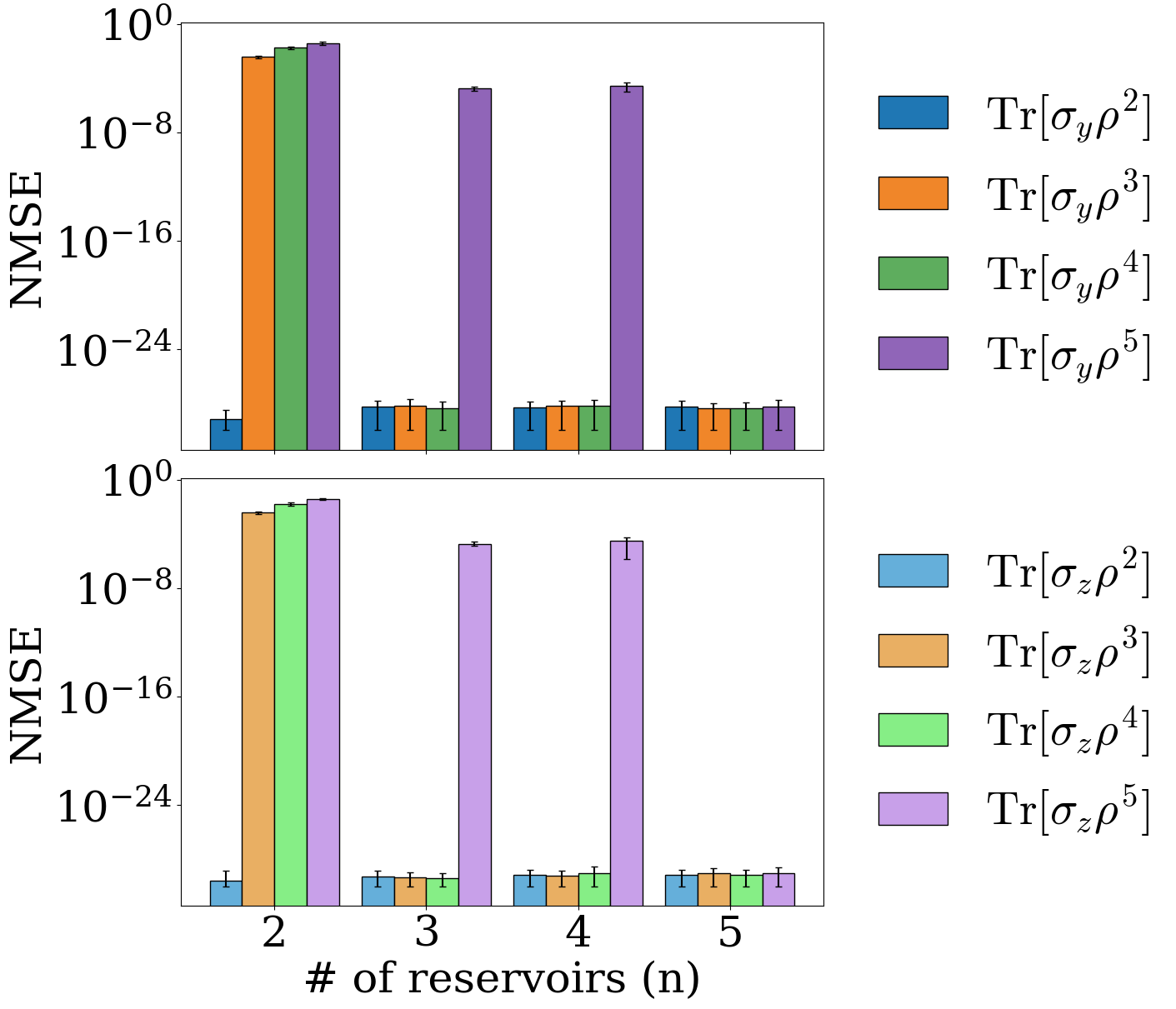}
    \vspace{-1mm}
    \caption{Reconstruction error in estimating polynomial targets of the form $Tr[O\rho^k]$ for a 1-qubit input, using the distributed architecture with a varying number of reservoirs. The top panel shows results for $O=\sigma_y$ and the bottom panel for $O=\sigma_z$, with $k=2,3,4,5$ in each case. The measurement basis, parameter choices, and statistical representation are the same as in Figure \ref{fig:smex}.}
    \label{fig:other}
\end{figure}

\begin{figure}[h!]
    \centering
    \includegraphics[width=0.9\linewidth]{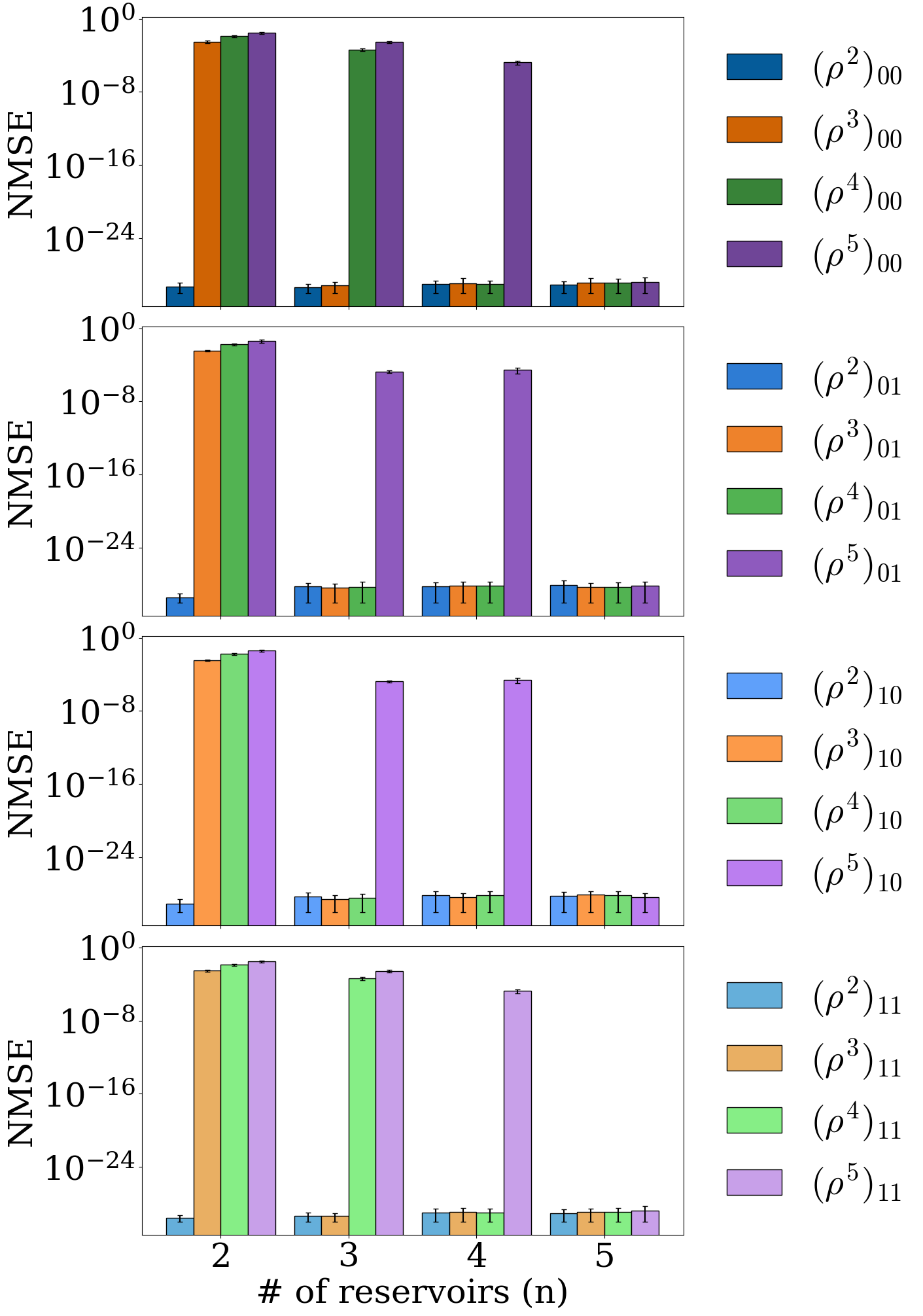}
    \vspace{-1mm}
    \caption{Reconstruction error in estimating the elements of $\rho^k$ for a 1-qubit input, using the distributed architecture with a varying number of reservoirs. The measurement basis, parameter choices, and statistical representation are the same as in Figure \ref{fig:smex}.}
    \label{fig:elements}
\end{figure}

The same analysis can be extended to other observables. For instance, Figure \ref{fig:other} reports the results for $\sigma_y$ and $\sigma_z$, which exhibit the same behavior as that previously discussed for $\sigma_x$. Alternatively, the element-wise reconstruction of $\rho^k$ can be considered, as shown in Figure \ref{fig:elements}. In this case, the populations share the same order of nonlinearity, and the coherences do as well. Notably, when combined in the trace, some of these nonlinearities may cancel out. This occurs, for example, in $\mathrm{Tr}[\rho^3]$, which requires only 2 reservoirs (Figure \ref{fig:full}), even though the individual populations of $\rho^3$ exhibit cubic complexity.

Overall, the key point is that with at least $k$ copies of the input, it is possible to reconstruct the entire symmetric subspace of $\rho^k$, and thus any observable that depends on it. Nonetheless, some specific information within this subspace may be reconstructible with fewer than $k$ injections, depending on the structure of the target.

\end{document}